\documentclass[sigplan,nonacm]{acmart}
\usepackage{popets}
\usepackage{cleveref}

\usepackage{xspace}

\usepackage{algorithm}
\usepackage{algorithmic}

\usepackage{xcolor}

\usepackage{newfloat}
\usepackage{listings}
\usepackage{graphicx,multirow,xspace,longtable}
\usepackage{tabularx,booktabs,blindtext,url}
\usepackage{makecell,amsmath,rotating}
\usepackage[font={small}, textfont=md]{caption}
\usepackage[T1]{fontenc}
\usepackage[utf8]{inputenc}
\usepackage[draft,inline,nomargin,index]{fixme}
\usepackage{marginnote}
\usepackage{tikz}
\usepackage{tcolorbox}
\usepackage{xcolor}
\usetikzlibrary{shapes, positioning, fit, backgrounds}

\setcopyright{none}
\crefformat{section}{\S#2#1#3}
\crefformat{subsection}{\S#2#1#3}
\crefformat{subsubsection}{\S#2#1#3}

\newcommand{\para}[1]{\noindent\textbf{#1}}
\newcommand{\parait}[1]{\textit{#1}}

\newcommand{\etc}{etc.}
\newcommand{\cf}{{\em cf.}\ }
\newcommand{\eg}{e.g.,\ }
\newcommand{\etal}{et al.\xspace}
\newcommand{\ie}{i.e.,\ }

\begin{document}

\title[]{On the Suitability of LLM-Driven Agents for\\Dark Pattern Audits}

\author{Chen Sun}
\orcid{}
\affiliation{%
  \institution{University of Iowa}
  \city{}
  \state{}
  \country{}}
\email{chen-sun@uiowa.edu}

\author{Yash Vekaria}
\orcid{}
\affiliation{%
  \institution{University of California, Davis}
  \city{}
  \state{}
  \country{}}
\email{yvekaria@ucdavis.edu}

\author{Rishab Nithyanand}
\orcid{}
\affiliation{%
  \institution{University of Iowa}
  \city{}
  \state{}
  \country{}}
\email{rishab-nithyanand@uiowa.edu}

\renewcommand{\shortauthors}{}

\begin{abstract}
  As LLM-driven agents begin to autonomously navigate the web, their ability to interpret and respond to manipulative interface design becomes critical. A fundamental question that emerges is: {\em can such agents reliably recognize patterns of friction, misdirection, and coercion in interface design (\ie dark patterns)?}
  We study this question in a setting where the workflows are consequential: website portals associated with the submission of CCPA-related data rights requests. 
  These portals operationalize statutory rights, but they are implemented as interactive interfaces whose design can be structured to facilitate, burden, or subtly discourage the exercise of those rights.
  We design and deploy an LLM-driven auditing agent capable of end-to-end traversal of rights-request workflows, structured evidence gathering, and classification of potential dark patterns.
  Across a set of 456 data broker websites, we evaluate: (1) the ability of the agent to consistently locate and complete request flows, (2) the reliability and reproducibility of its dark pattern classifications, and (3) the conditions under which it fails or produces poor judgments.
  Our findings characterize both the feasibility and the limitations of using LLM-driven agents for scalable dark pattern auditing.

\end{abstract}

\keywords{Dark patterns, audit, LLM, agent, CCPA.}

\settopmatter{printacmref=false}
\renewcommand\footnotetextcopyrightpermission[1]{}
\maketitle

\section{Introduction}  \label{sec:introduction}
Dark patterns are interfaces that subvert, impair, or distort user decision-making. 
These interfaces have been documented across domains including e-commerce \cite{mathur2019dark}, consent interfaces \cite{Utz-CCS2019}, and subscription flows \cite{Gray-CHI2025}.
Early work by Gray \etal \cite{gray2018dark} characterized recurring manipulative design strategies in user interfaces, while large-scale measurement by Mathur \etal \cite{mathur2019dark} identified widespread deceptive interfaces across thousands of shopping websites.
Experimental evidence has also shown that such designs can influence user decisions \cite{luguri2021shining}.
In consent contexts specifically, studies have shown that cookie banners and consent dialogs frequently deploy asymmetric choice architectures that bias user behavior \cite{Utz-CCS2019}.
Despite these advances, dark pattern audits remain largely manual.
Researchers typically combine automated crawling with extensive human review to interpret interface elements relative to normative taxonomies \cite{mathur2019dark,nouwens2020dark}.
This approach has high fidelity but is difficult to scale across heterogeneous interaction flows occurring on a large sample of websites and is challenging to reproduce when classification requires contextual interpretation.
In other words, while dark patterns have been extensively documented, the process of auditing them remains labor-intensive, difficult to replicate, and constrained in scale.

Simultaneously, a parallel line of research has demonstrated that LLM-driven agents can navigate complex web environments using semantic reasoning rather than brittle site-specific scripts.
WebGPT \cite{nakano2021webgpt} showed that LLMs can interact with web pages to retrieve and synthesize information. 
ReAct \cite{yao2022react} introduced a framework for combining  reasoning and action to enable structured decision-making in interactive environments.
WebArena \cite{zhou2023webarena} and Mind2Web \cite{deng2023mind2web} further demonstrate that LLMs are able to complete realistic interaction tasks across a wide variety of interfaces.
These capabilities of LLM-driven agents raise a natural question: {\em can LLM-driven agents  serve as instruments for dark pattern auditing?}
Unfortunately, this is not answered by the current literature which is focused primarily on assessing the task completion capabilities of LLM-driven agents and not on whether agents can assess characteristics of interface design, distinguish standard from manipulative workflows, or generate reproducible, evidence-backed normative judgments.
Put differently, the question that remains unanswered is not whether LLM-driven agents can click or submit forms, but whether they can reliably characterize the structure of available interaction flows on a website to a predefined dark pattern taxonomy. 

The California Consumer Privacy Act (CCPA) data rights request portals provide a particularly consequential setting to examine whether LLM-driven agents can reliably recognize dark patterns in interfaces. 
The CCPA established statutory rights including access, deletion, and the right to opt out of the sale or sharing of personal information.
These rights are operationalized through web-based submission workflows implemented independently by companies that are subject to the regulation.
Because exercising these rights requires complex navigation of websites and interactive forms with many branching possibilities and verification steps, the design of these interfaces can either facilitate or impede user agency.
While prior work has examined compliance and usability in consent and GDPR contexts \cite{Utz-CCS2019,nouwens2020dark,toth2022dark,santos2026usable}, large-scale empirical evaluation of manipulative design remains limited, as do audits of dark patterns within CCPA data rights submission flows.

Taken together, we lack an understanding of whether LLM-driven agents are suitable for dark pattern auditing in complex, multi-step workflows such as CCPA rights-request portals.
Specifically, it is unclear whether such agents can: (1) consistently traverse heterogeneous interfaces, (2) identify workflows containing characteristics of well-known dark patterns; and (3) produce evidence-based, reproducible classifications of dark patterns tied to a predefined taxonomy.
These questions concern not only the capability of LLM-driven agents, but also their reliability and limitations.
Filling this research gap is critical because a principled evaluation of LLM-driven dark pattern auditing is needed to determine whether regulatory interfaces, where policy regulates user experience, can be monitored with consistency and evidentiary rigor at scale.
Without such an evaluation, audits of CCPA rights-request portals and similar systems will remain episodic, resource-intensive, and difficult to reproduce and scale.
Conversely, premature reliance on unvalidated automated agents risks generate misleading conclusions about the prevalence and nature of dark patterns.
In both cases, the enforceability of statutory rights is weakened if the interfaces through which those rights are exercised cannot be systematically examined.

The overall objective of this work is to evaluate the feasibility, reliability, and limitations of LLM-driven agents for identifying evidence-backed dark patterns in multi-step web interaction flows, using CCPA data rights request portals as a testbed.
We provide an empirical characterization of both the capabilities and failure modes of these agents, thereby clarifying when agentic dark pattern auditing is appropriate and when it is not.
We achieve this objective by answering the following research questions.

\parait{RQ1. Can LLM-driven agents reliably audit dark patterns?}
We implement an instrumented LLM-driven browser agent capable of end-to-end traversal of CCPA Right-to-Access workflows, structured evidence collection, and taxonomy-based dark pattern classification. 
We construct a human-annotated ground-truth dataset across 100 data brokers and evaluate multiple prompting strategies through a controlled ablation study. 
We measure execution robustness, classification accuracy, explanation validity, and pattern-level performance. 
Under the best-performing configuration, our implementation achieves strong detection performance and high explanation accuracy for several obstruction-oriented patterns. 
However, execution instability and security barriers reduce effective task completion, indicating that reliability depends not only on reasoning quality but also on the robustness of browser interaction.

\parait{RQ2. What limits constrain LLM-driven agents during dark pattern audits?}
We conduct a structured failure analysis of our implementation across observation, reasoning, and action stages of the agent's interactions. 
We identify recurring perceptual blind spots (\eg limited encoding of visual salience), reasoning errors (\eg speculative navigation), and execution-level failures (\eg CAPTCHA barriers and incomplete form interaction). 
These findings characterize the observed boundary conditions of our agent-driven auditing pipeline in the CCPA context. 
While alternative architectures or training regimes may mitigate some of these limitations, our results demonstrate that autonomous dark pattern auditing, as implemented and evaluated here, remains constrained and benefits from human oversight in regulatory audit settings.

\section{Conceptual and Methodological Background}
\label{sec:conceptual}

This section establishes the conceptual and methodological foundation for our study.
Our goal is not merely to apply an LLM agent to a new domain, but to evaluate whether LLM-driven agents can serve as instruments for interaction-level dark pattern auditing.
To do so, we first describe how dark pattern detection has traditionally been framed, then define interaction-level dark pattern auditing as a measurement problem, and finally explain the suitability of the CCPA's right-to-access portal interfaces as a testbed for evaluating the capabilities and limitations of autonomous dark pattern auditing.
The design described here structures the remainder of the paper.
In \Cref{sec:design}, we describe the experimental design used for our audit.
\Cref{sec:rq1} evaluates the LLM-driven agent's effectiveness under this design. 
\Cref{sec:rq2} then characterizes various failures of the agent.

\subsection{Towards dynamic interaction-level auditing}
\label{sec:conceptual:moving}
Early large-scale detection approaches relied on HTML crawling and manual validation to identify deceptive elements embedded in page content \cite{mathur2019dark}.
Subsequent systems incorporated advancements in computer vision and natural language processing techniques to analyze web and mobile UI screenshots \cite{mansur2023aidui, chen2023unveiling}.
More recently, multi-modal LLMs have been used to detect deceptive design patterns from static UI artifacts and generate structured explanations for their classifications \cite{kocyigit2025deceptilensc}.
These approaches represent significant progress in automating detection.
However, they primarily operate on static representations of user interfaces such as screenshots, HTML segments, or isolated UI states.
As prior work acknowledges, certain deceptive patterns cannot be reliably assessed from screenshots alone, particularly when manipulation emerges only through user interaction, multi-step navigation, or contextual transitions \cite{kocyigit2025deceptilensc}.
Further, empirical studies of opt-out or right-to-access workflows in the context of the CCPA and GDPR show that friction may arise from cross-page inconsistencies, fragmented instructions, and pathological verification flows \cite{habib2019empirical, pohn2023needle, lobel2024access}.
These findings motivate a shift in framing: from static to dynamic interaction-level auditing.

With this framing, we define interaction-level dark pattern auditing as the process of: 
(1) traversing a multi-step workflow from end to end;
(2) recording structured evidence of interface behavior, navigation paths, and system responses; and
(3) mapping the resulting interaction trace to a predefined dark pattern taxonomy.
The interaction trace includes navigation history, branching logic, required form fields, verification requirements, among other observable information.
A valid audit requires trace-linked evidence such as URLs, interface text, and other details required to complete independent manual verification.
This framing treats dark pattern auditing as a problem of systematically instrumenting and evaluating interaction workflows, rather than classifying isolated interface artifacts.
However, it brings new challenges: we are no longer tasked with simply recognizing deceptive elements, but doing so consistently across heterogeneous, dynamic workflows while preserving evidentiary rigor and reproducibility.
Although manual audits can already achieve this along with more interpretive depth, they are labor-intensive, difficult to standardize, and challenging to replicate at scale.
The central question is therefore whether LLM-driven agents can improve the feasibility of such audits.

\subsection{Normative operationalization} 
\label{sec:conceptual:taxonomy}
Dark patterns are not purely technical features of an interface or workflow; they involve normative judgments about whether a design unfairly influences or impairs user decision-making.
Prior work has identified recurring manipulative strategies across domains \cite{gray2018dark} and, more recently, organized these strategies into structured ontologies to provide shared terminology and conceptual clarity \cite{gray2024ontology}.
Regulatory bodies have also articulated principles for classifying manipulative design, emphasizing clarity, symmetry of choice, and avoidance of practices that impair decision-making \cite{FTC2022dark, oecd2022dark, edpb2023darkpatterns}.
These principles are particularly relevant in privacy rights workflows, where interface design directly affects the exercising of legally protected rights.
Our work builds on this established foundation rather than introducing new definitions or identifying new patterns.
To operationalize these ideas in our auditing framework, we apply two explicit criteria.
First, we use a {\em consumer expectation test}, asking whether a given interface behavior violates reasonable expectations regarding clarity, symmetry, or ease of execution.
Second, we apply {\em harm mechanism mapping}, identifying the specific way in which a design may create friction, confusion, or obstruction.
Importantly, applying these criteria does not eliminate interpretive ambiguity.
In regulatory contexts, some behaviors such as identity verification requirements may be necessary safeguards in one setting and disproportionate barriers in another.
Rather than treating these cases as simple annotation errors, we treat this ambiguity as part of what must be evaluated.
Our goal is not to assume that all borderline cases have a clear binary answer, but to assess how reliably an LLM-driven auditing agent can operate within these structured yet inherently normative constraints.

\subsection{CCPA portals as a testbed}
\label{sec:conceptual:portals}
We evaluate interaction-level auditing within the context of the CCPA, which grants consumers rights to access, delete, and opt-out of the sharing of personal information \cite{CCPA}.
Businesses subject to the CCPA must provide designated mechanisms for exercising these rights and avoid interface practices that impair consumer decision-making.
Our study focuses exclusively on data brokers who have self-registered with the California Privacy Protection Agency (CPPA) as being subject to the CCPA.
This data broker registry \cite{databrokers-cppa} provides a population of entities that explicitly acknowledge their obligations under the CCPA, thus allowing us to circumvent the challenging problem of identifying websites that are subject to the CCPA \cite{van2022setting, musa2024c3pa}.
The right-to-access workflows made available within this population of data brokers present a suitable testbed for our study for three reasons. 
First, they are structurally complex. 
As shown in prior work \cite{pohn2023needle, lobel2024access}, exercising data rights typically requires multi-step navigation for locating privacy disclosures, identifying valid channels for submitting requests, completing forms, and locating identity verification requirements.
%
%
Second, these workflows embed both legitimate safeguards and potential manipulation.
In particular, identity verification and submission constraints may be needed for compliance or security, while at the same time can be structured in ways that introduce unnecessary friction or obscure navigation paths.
Distinguishing between legitimate security measures and dark patterns is therefore challenging and context-dependent.
Finally, unlike marketing consent banners or promotional flows, CCPA right-to-access workflows directly mediate legally protected consumer rights and errors in classification have implications for regulatory enforcement.

\section{Experiment Design and Methodology} \label{sec:design}

\begin{figure*}
    \centering
    \includegraphics[width=1\linewidth]{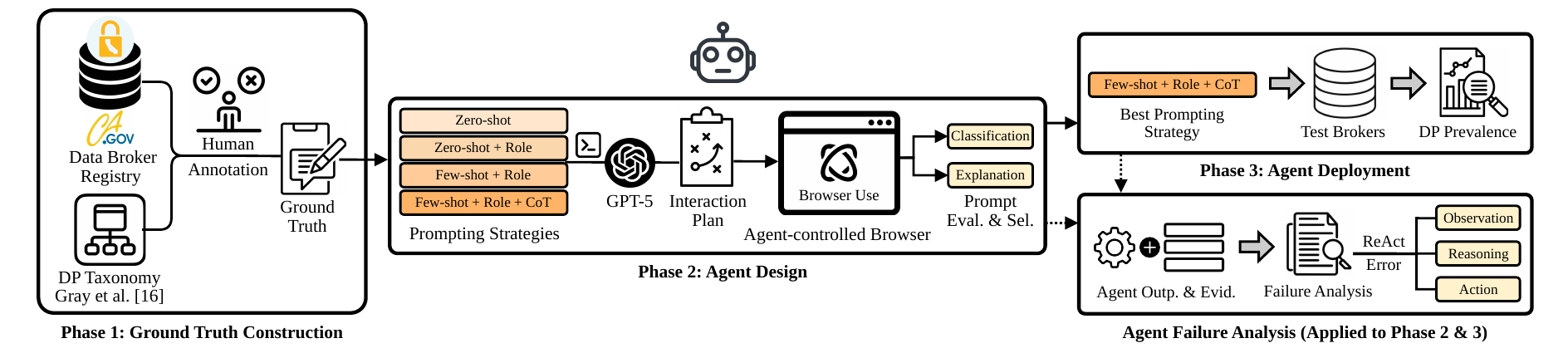}
    \caption{Overview of our experiment design and methodology (\cf \Cref{sec:design}).}
    \label{fig:methodology}
    \Description{A multi-phase workflow diagram illustrating the study design.}
\end{figure*}

Here, we describe the methodology used to evaluate LLM-driven agents for dark pattern auditing in the context of CCPA's right-to-access workflows.
Our study consists of three phases. These are illustrated in \Cref{fig:methodology}.
First, we construct a human-annotated ground truth dataset through direct exercising of the right-to-access workflows on a subset of data brokers subject to the CCPA (\Cref{sec:design:groundtruth}).
Next, we design an LLM-driven browser agent and conduct a controlled prompt ablation study on the annotated subset to evaluate the reliability and robustness of dark pattern detection (\Cref{sec:design:ablation}).
Finally, we deploy the most performant configuration at scale across the remaining data brokers to derive an estimate of the prevalence and distribution of dark patterns in right-to-access workflows, as well as to assess agent robustness in large-scale deployments (\Cref{sec:design:deployment}).

\subsection{Study design and dataset construction} \label{sec:design:dataset}

Our analysis focuses on data brokers that are self-registered with the California Privacy Protection Agency (CPPA) Data Broker Registry \cite{Data_Brokers_Registry}.
From this registry, we extracted 475 unique website links and used Playwright \cite{Playwright} to verify website accessibility.
We removed 19 websites which were unreachable, resulting in a final dataset of 456 unique websites associated with data brokers that were subject to the CCPA.
All experiments described in the remainder of this paper are conducted on this set.
We intentionally analyze the right-to-access process as a multi-step workflow rather than relying on isolated UI screenshots.
As noted in prior work \cite{kocyigit2025deceptilensc, Gray-CHI2025} and in our own observations, several dark pattern categories (\eg Privacy Mazes and Feedforward Ambiguity) manifest only through specific multi-step interactions, navigation sequences, or conditional page transitions.
Static screenshots are unable to capture these behaviors and their contextual appropriateness.
Therefore, both the human annotation process and the agent evaluation replicate the same stepwise right-to-access procedures, while recording complete interaction traces.
We adopt the three-level dark pattern ontology proposed by Gray \etal \cite{gray2024ontology} as our seed taxonomy. From this ontology, we focused on the patterns that manifested during the manual exercising of the right-to-access workflows.

\subsection{Phase 1: Ground truth construction} \label{sec:design:groundtruth}
In this phase, our goal is to manually examine the right-to-access workflows associated with a random sample of 100 websites in our dataset and construct a human-annotated dataset of dark patterns based on this examination. 
This dataset serves two purposes.
First, it establishes an empirically derived benchmark for evaluating the LLM-driven auditing agent in Phase 2.
Second, it defines the interaction protocol that the agent will later attempt to replicate.

\para{Interaction protocol for exercising right-to-access workflows.}
For each of the 100 randomly selected data broker websites, one annotator, who was located in California to avoid potential geo-blocks or content variations, completed a right-to-know workflow interaction protocol using their own personal data.
For each broker, the interaction protocol involved the following steps: 
(1) visiting the website linked in the CPPA's Data Broker Registry \cite{databrokers-cppa}, 
(2) locating the privacy policy or other pages that might contain information related to CCPA-granted data rights, 
(3) reviewing the identified pages to locate information about how users might exercise their right-to-access, including available submission mechanisms and verification requirements, 
(4) interacting with any online right-to-access request submission mechanisms (\eg webforms) to reveal required input fields, available options, conditional interfaces, and identity verification requirements.
The annotator did not submit the final request --- \ie no physical or digital right-to-access request submissions were completed. 
This decision was made to avoid imposing unnecessary burden on brokers while allowing full observation of all pre-submission interactions.
To preserve reproducibility and evidentiary rigor, each interaction performed by the annotator was comprehensively documented. 
We recorded full-session screen captures, logged full-page screenshots at each navigation stage, and maintained detailed annotator notes describing the observed interface elements, required actions, and navigation paths.  
To ensure low friction for logging annotator observations, the annotator was told to provide audio commentary of their observations during interactions with each right-to-access workflow.
By standardizing the interaction sequence and documentation process across all the sampled 100 brokers, we are able to minimize procedural variance to the best extent possible and ensure that differences in annotator classifications reflect site-level design characteristics.
This standardized interaction protocol defines the reference auditing procedure against which the LLM-driven agent is evaluated in Phase 2.

\para{Review of dark pattern literature.}
Two annotators first reviewed the dark pattern literature and regulatory guidance to establish a shared understanding of each dark pattern type.
Specifically, we relied on the ontology by Gray \etal \cite{gray2024ontology} to define categories; on CCPA guidance \cite{CCPA}, EDPB guidelines \cite{edpb2023darkpatterns}, FTC reports \cite{FTC2022dark},  OECD reports \cite{oecd2022dark}, and prior research  \cite{gray2018dark, tran2025dark} for concrete examples of dark patterns and for clarifying what constitutes reasonable design in the context of exercising a right-to-access workflow.

\para{Annotation of dark patterns.}
For each of the 100 data brokers in the ground truth sample, annotators reviewed the complete interaction trace, including full-session recordings, page-level screenshots, and detailed annotator notes.
Each candidate dark pattern category was then evaluated using a two-step assessment.

\parait{(1) Reasonable expectation assessment.} The annotators asked whether the interface violated reasonable expectations of clarity and simplicity of design, presentation, and instructions, as implied by CCPA's requirement that rights mechanisms be clear, not misleading, and not impairing consumer choice.
This included assessing whether submission pathways were discoverable without excessive navigation, instructions were consistent, any required steps appeared necessary/reasonable, and interfaces conveyed the actions they enabled.
    
\parait{(2) Harm mechanism assessment.} If a violation of the reasonable expectation test occurred, the annotator performed an assessment of the harm mechanism --- \ie the category assignment was determined by the primary mechanism through which the workflow impaired the rights exercise process, not by the mere presence of friction.
When multiple distinct harm mechanisms were observed in the same workflow, multiple categories were assigned.
For example, consider the workflow in which the privacy policy states that consumers may submit a Right-to-Access request ``via email or webform'', but lists no email address.
The only working submission mechanism is a link labeled ``Do Not Sell My Personal Information''. 
This single workflow may receive multiple dark pattern assignments: 
First, the contradiction between the information presented in the privacy policy and the available submission mechanism would be mapped to {\em conflicting information} or {\em hidden information}, depending on whether the email address is located in other sections of the document or nowhere.
Second, the misleading link label (``Do Not Sell My Personal Information'') for the right-to-access request mechanism would be mapped to {\em information without context} or {\em feed-forward ambiguity}, depending on whether the interface clarifies that the same workflow also handles right-to-access requests or not.

\para{Assessment of annotator reliability.}
To assess annotation reliability, we conducted a two-stage process\footnote{The annotation guide along with the annotated data for the 100-broker ground truth sample of brokers is available at: \url{https://osf.io/8ab4g/overview?view_only=685294a249b44d8bbdf078646c3615b4}.}.
First, the annotators independently reviewed an initial subset of 20 data brokers using the same protocols and interaction traces.
For each dark pattern annotation, the annotators recorded the specific navigation path, the exact UI text or interface element involved, supporting HTML evidence if applicable, and a written rationale explaining both (1) how the interface violated the reasonable expectations assessment and (2) how the observed behavior mapped to the annotated dark pattern category.
The inter-annotator agreement was then measured using Cohen's $\kappa$ on binary (present/absent) labels for each dark pattern category, yielding $\kappa = 51.9\%$.
Next, the annotators discussed the source of their disagreements and clarified boundary cases.
Following this, they independently annotated an additional 20 data brokers. 
Agreement on this second subset increased to $\kappa = 73.7\%$, indicating substantial agreement, and hence, increased consistency.
Any disagreements within these 40 jointly reviewed data brokers were resolved through discussion.
After this calibration phase, one of the annotators applied the refined annotation protocol consistently to the remaining 60 data brokers.
Across the 100 data brokers, the annotators identified 14 recurring dark pattern categories from the Gray \etal ontology \cite{gray2024ontology}.
Importantly, the 40 jointly annotated workflows included instances of all fourteen categories retained for evaluation, ensuring calibration before completing the remaining annotations.
However, six of these appeared fewer than 20 times in the complete 100-sample dataset.
Because such small counts do not permit statistical rigor during the evaluation of our agent, potentially causing performance within a few cases to disproportionately influence our understanding of the agent's general performance for that category, they were excluded from our evaluation experiments.
The remaining eight categories were retained for systematic evaluation in Phase 2.
In ~\Cref{tab:dp_harm_mechanism}, we present the definitions of the eight high-frequency dark patterns and one representative example of a harm mechanism observed in the 100 data broker annotation workflows (\cf \Cref{sec:appendix:harm_mapping} for additional examples).

\begin{table*}[t]
\caption{High-frequency dark pattern categories in right-to-access workflows observed in our ground truth sample of 100 data brokers.}
\label{tab:dp_harm_mechanism}
\centering
\small

\resizebox{\textwidth}{!}{%
\begin{tabular}{l p{6.8cm} p{7.5cm}}
\toprule[1.2pt]
\bfseries Pattern & \bfseries Definition & \bfseries Harm Mechanism Example \\
\midrule

Adding Steps 
& Subverts the user’s expectation that a task will take as few steps as technologically necessary, instead introducing additional required interactions. 
& One identifier per submission forces users with multiple identifiers (e.g., email address, IP address, MAID) to repeat the full request process multiple times. \\
\addlinespace
Conflicting Info. 
& Includes two or more sources of information that conflict with each other, creating uncertainty about the consequences of the user’s actions. 
& Different pages list incompatible official submission methods (e.g., one page states email is valid, while another states only a webform is accepted). \\
\addlinespace
Creating Barriers 
& Subverts the user’s expectation that relevant tasks will be supported by the interface, instead preventing or unnecessarily complicating the task to disincentivize user action. 
& The request requires highly sensitive or burdensome materials (e.g., government-issued ID, SSN, biometric verification) that exceed what is reasonably necessary. \\
\addlinespace
Feedforward Ambiguity 
& Subverts the user’s expectation that their choice will result in a predictable outcome, instead creating a discrepancy between provided information and the resulting action. 
& A link promises a general privacy or access request pathway, but the destination interface supports only a limited subset of rights or redirects users into an unrelated workflow (e.g., account creation). \\
\addlinespace
Hidden Info. 
& Subverts the user’s expectation that relevant information will be accessible and visible, instead disguising it or framing it as irrelevant. 
& A section framed as comprehensive (e.g., “California Privacy Rights”) omits key submission methods that exist elsewhere without cross-reference, leading users to stop searching prematurely. \\
\addlinespace
Info. Without Context 
& Alters relevant information or user controls in ways that limit discoverability, making it unlikely that users will find the information or action they seek. 
& Access submission instructions appear only in unrelated sections (e.g., “Contact Us” or “Opt-Out”) and are not referenced in the expected rights section. \\
\addlinespace
Privacy Mazes 
& Requires users to navigate through multiple screens or fragmented pathways to accomplish a task, without providing a clear or comprehensive overview. 
& Submission methods are distributed across multiple pages, each containing partial guidance, and no single page presents a complete set of instructions. \\
\addlinespace
Visual Prominence 
& Places an element relevant to user goals in visual competition with a more distracting and prominent element, thereby diverting attention from the user’s intended task. 
& Non-essential elements visually dominate or compete with privacy controls, diverting users from completing their request. \\

\bottomrule[1.2pt]
\end{tabular}%
}
\end{table*}

\para{Limitations.}
This ground truth dataset is limited to pre-submission right-to-access workflows.
We do not evaluate post-submission compliance behavior, responses, or verification practices, which have been shown to introduce additional user burdens in recent research \cite{van2025consumer}.
Additionally, the right-to-access workflows were identified and executed by a single user and may not capture all variations.
Further, although annotation decisions were grounded in regulatory guidance and validated taxonomies, the identification of dark patterns still remains interpretive.
Finally, dark pattern categories observed fewer than 20 times were excluded from evaluation to preserve statistical validity, which limits the systematic evaluation of rare dark patterns.
Despite these limitations, this dataset provides a standardized, evidence-backed benchmark for evaluating agent performance in interaction-level dark pattern auditing. 

\subsection{Phase 2: Agent design} \label{sec:design:ablation}
Here, we evaluate whether an LLM-driven browser agent can reproduce the standardized right-to-access interaction protocols defined in Phase 1 and correctly identify dark patterns relative to the human-annotated benchmark.
This phase isolates the interaction-level dark pattern identification problem under controlled conditions before scaling to the full dataset of 456 data brokers in Phase 3.

\para{Agent interaction and logging protocols.}
We implement our auditing agent using the browser-use framework (v0.9.5) \cite{browseruse2025}.
The agent interacts with live websites using Playwright and derives planning, reasoning, and vision capabilities used to interpret HTML/DOM structures and screenshots from the GPT-5 LLM.
For each of the 100 data brokers in the ground truth dataset, the agent is instructed to perform the same interaction sequence as the Phase 1 manual examination of right-to-access workflows.
To ensure that the agent's decisions can be audited, we record the full interaction trace, including full-page screenshots and HTML/DOM structures captured by the agent, session video recordings, extracted page content, the agent's internal reasoning steps, and task completion metadata, including intermediate task completion status, token usage, and duration.
For each broker, the agent outputs a Pydantic structured JSON object containing: 
(1) the detected right-to-access request submission mechanisms;
(2) the complete list of webform fields (if applicable);
(3) a binary (present/absent) label for each of the eight dark pattern categories identified in Phase 1; and
(4) for each identified dark pattern: the specific navigation paths, interface elements, and actions that surface the pattern, along with a detailed explanation for the decision (based on the reasonable expectation and harm mechanism assessments) and a confidence score.
This structured output ensures that each classification can be traced back to observable evidence.
The agent also logs any failures or issues encountered during automation, such as unresponsive UI elements or CAPTCHA challenges that prevent access.

\para{Prompting configurations.}
Because the performance of our agent is highly dependent on how the task is defined in the input prompt, we systematically evaluate how guidance in the prompt affects dark pattern detection using an ablation design \cite{singhal2023large}.
The full prompting configurations are illustrated in Appendix~\Cref{sec:appendix:prompt}, \Cref{fig:prompt_1,fig:prompt_2,fig:prompt_3}.

\parait{Level 1: Zero-shot prompting.} 
This prompting strategy establishes the baseline performance. 
Here, the prompt only contains the task description which instructs the agent to follow the predefined manual right-to-access workflow from Phase 1, the formal definitions of the eight dark pattern categories identified in Phase 1, and the required JSON output schema.
Thus, the agent relies entirely on it's underlying pretrained model, without additional context, to complete the interaction-level audit.

\parait{Level 2: Zero-shot prompting with a defined role.}
We augment the baseline prompt with explicit role conditioning, framing the model as an auditor responsible for evaluating CCPA right-to-access mechanisms.
Role-based prompting has been shown to meaningfully affect model performance on specialized tasks, even without providing specific examples \cite{reynolds2021prompt}.
This configuration tests whether the role improves the model's alignment with regulatory criteria (\eg the reasonable expectation assessment).

\parait{Level 3: Few-shot prompting with a defined role.}
In this configuration, we retain the regulatory role and introduce scenario-based few-shot examples derived from our ground truth dataset.
To construct these examples, two authors systematically reviewed the annotated workflows and clustered instances within each dark pattern category based on their subtype (\ie each low-level pattern within a meso-level pattern is a subtype \cite{gray2024ontology}).
Then, for each subtype, we constructed one canonical scenario that abstracted away broker-level details while preserving the structural harm mechanism.
Each few-shot scenario includes: a description of the problematic interface, the violated consumer expectation, and the resulting harm to the right-to-access workflow.
To ensure broader conceptual coverage, we also reviewed prior literature and regulatory reports to identify subtypes of a pattern that were not represented in our dataset and created abstract canonical examples for these.
The agent is then instructed to use these canonical examples for each dark pattern as references when mapping their observed behaviors to dark patterns.
This operationalizes standard few-shot in-context learning \cite{brown2020language} for interaction-level dark pattern audits in right-to-access workflows.

\parait{Level 4: Few-shot prompting with defined role and chain-of-thought reasoning.}
The final configuration augments the few-shot strategy with explicit chain-of-thought reasoning.
For each scenario, we include: the raw UI element that triggers concern; the description of the observed interface behavior; the specific actions taken to reveal the dark pattern; the resulting dynamic outcome; and a step-by-step rationale which explains what the user encounters, why it impairs their ability to exercise the right-to-access, and why it satisfies the dark pattern definition.
The agent is instructed to follow this same reasoning structure in its outputs.
This configuration tests whether structured intermediate reasoning improves interaction-level auditing capabilities and explanation quality, consistent with findings that chain-of-thought reasoning can improve performance on multi-step tasks \cite{wei2022chain}.

\para{Failure reporting and workflow completion verification.}
The browser-use framework provides an internal flag, \texttt{is\_successful},  indicating that the agent has completed its planned interaction sequence.
However, as observed during our pilot testing, this internal signal can indicate success even when the agent has actually failed to complete the right-to-access workflow.
For example, the agent may terminate after partial interaction with a form or without fully exposing required fields (\eg by not hitting the `{\em next page}' button on a form).
To obtain a more accurate measure of workflow completion, we introduce a structured verification procedure. 
Initially, for each prompting strategy, if the agent failed to complete the right-to-access workflow, it was initially instructed to provide a description of the issues it encountered in free-text form. 
We manually reviewed these descriptions and grouped them into six recurring failure categories. 
Following this categorization, we required the agent to explicitly classify failure-causing issues into one of these six predefined categories and provide supporting evidence.
Runs containing such failure reports are treated as unsuccessful, even if the internal {\tt is\_successful} flag indicates completion.

\para{Failure categorization.}
These six categories, defined below, capture distinct ways in which agent execution of a workflow can fail.

\parait{(1) Automation instability.} This category refers to failures in the automation infrastructure rather than the website or the agent's own reasoning. 
These include unexpected browser crashes, network instability, and execution timeouts that prevent the agent from completing the workflow.

\parait{(2) Agent instability.} This label is assigned when the agent is unable to complete a workflow due to internal failures in execution or planning.
This includes cases where the intermediate documents required for planning (\eg the structured JSON outputs and reasoning traces) are malformed, incomplete, or inconsistent with the expected schema.
In these cases, the failure arises from the instability in the agent's internal state rather than issues with the automation infrastructure or website/workflow artifacts.

\parait{(3) Security barriers.} This label is assigned to failures caused by security mechanisms deployed on websites that hinder agent interaction.
These include CAPTCHA challenges, bot-detection systems, or rate-limiting mechanisms that block the agent from interacting with the website.

\parait{(4) Content format limitations.} This category refers to cases in which relevant instructions or submission details are embedded in formats that cannot be programmatically accessed or interacted with by an agent, including PDFs whose text cannot be extracted or embedded hyperlinks that cannot be traversed by the agent's browser.
In these cases, the right-to-access workflows may exist, but are not described in a way that can be used by our agent.

\parait{(5) Navigation failure.} This label is assigned when the agent follows the interaction protocol but is unable to locate a right-to-access workflow.
This may occur when links are broken, disclosures are not accessible, or mechanisms are not discoverable through expected navigation.
In some cases, such lack of discoverability could itself reflect a dark pattern, such as hidden information or a privacy maze.
However, in this study we conservatively classify these cases as execution failures rather than dark patterns. 
We adopt this approach because, without successfully reaching the relevant submission pathway, the agent cannot reliably determine whether the failure is due to a dark pattern or underspecified interaction protocol.

\parait{(6) Interaction failure.} This label is assigned when the agent successfully reaches a right-to-access submission mechanism but fails to fully expose required input fields or conditional verification steps due to incomplete interaction with dynamic interface elements (\eg failing to trigger a multi-page form). 
In these cases, the pathway is located, but the workflow is not fully revealed.

\para{Prompt evaluation and selection.}
We evaluate each prompting strategy by comparing the agent's outputs against the Phase 1 human-annotated ground truth dataset.
We use two metrics.
The first is {\em classification accuracy}: for each broker-pattern pair, we evaluate whether the agent correctly identifies the pattern as present or absent, treating the task as binary classification per category.
The second is {\em explanation accuracy}: for each correctly classified instance, we verify whether the agent's rationale references observable evidence in the recorded interaction trace and correctly maps that evidence to the correct predefined harm mechanism.
An explanation is considered accurate only when both the decision and supporting evidence align with the ground truth rationale.
We select the final prompting configuration by jointly considering the classification and explanation accuracy, thus prioritizing correctness, as well as interpretable and verifiable reasoning.

\para{Separation of accuracy and failure analysis.} 
In the remainder of this paper, runs classified as unsuccessful are excluded from any analysis of the agent's dark pattern detection performance.
This separation ensures that any reported measures of dark pattern detection and explanation accuracy reflect classification performance within completed workflows rather than failure rates.

\para{Limitations.}
Phase 2 only evaluates the agent on the 100 data broker right-to-access workflows contained in our ground truth dataset.
This controlled setting allows direct comparisons to human-annotated ground truth, but also limits evaluation to the types of workflows and dark pattern subtypes observed within that dataset.
Although we expand the few-shot examples using prior literature to improve conceptual coverage, rare or previously unseen pattern variants may not be fully represented.
The evaluation also assumes a fixed interaction protocol (defined in Phase 1) and does not consider alternative navigation strategies that different users might follow.
Further, the workflow completion verification procedures improve our ability to measure workflow execution success, but relies on structured self-reports from the agent, which may by itself contain errors.
However, our pilot studies show that these errors are rare and that our failure categories cover all observed agent failures.
Taken together, these limitations mean that our results pertaining to the agent's interaction-level dark pattern detection performance are obtained under controlled standardized conditions rather than across all possible real-world variations.
Despite these limitations, our approach is able to isolate the agent's interaction-level dark pattern identification capabilities, enabling a principled comparison across different prompting strategies.

\subsection{Phase 3: Agent deployment} \label{sec:design:deployment}
Phase 3 extends our evaluation beyond the 100-broker ground truth dataset to the remaining 356 brokers. 
While Phase 2 isolates agent performance under controlled benchmark conditions, Phase 3 examines how the agent behaves under real-world variability. 

\para{Large-scale deployment protocol.}
Using the most performant prompting strategy identified in Phase 2, we deployed the agent across the remaining 356 data brokers that were not part of our ground truth dataset.
The model version, browser-use framework, input and output schemes, interaction trace logging pipelines, and evidence logging pipelines remain unchanged from Phase 2.

\para{Limitations.}
Phase 3 evaluates the agent at scale across the remaining 356 data broker right-to-access workflows, but it does not include human-annotated ground truth for these workflows.
Therefore, we primarily use the results to assess agent failure incidence rates in large-scale deployments and to derive an estimate of the prevalence and distribution of dark patterns observed in current right-to-access workflows, rather than correctness of this estimate.

\subsection{Ethics considerations}\label{sec:design:ethics}
This study involves automated interaction with publicly accessible right-to-access workflows on data broker websites. 
All interactions were limited to pre-submission steps. 
Neither the agent nor the annotator submitted requests, transmitted personal data, or retrieved consumer records. Interaction rates were kept low to avoid imposing operational burden.
When workflows were blocked by CAPTCHA or bot-detection systems, we treated these cases as execution failures and did not attempt to circumvent protective mechanisms. 
We deliberately avoided evasive crawling techniques (\eg disguised user agents or proxy rotation), recognizing that such approaches may violate deployment norms and undermine website safeguards. 
This imposes a principled limit on coverage.
Dark pattern identification inherently involves interpretive judgment, even when made by autonomous agents. 
We therefore frame our results as descriptive measurements of interface structure rather than determinations of legal non-compliance.

\begin{table*}[t]
\caption{
{\it RQ1: Performance across four prompting strategies (\cf \Cref{sec:rq1:ablation}).}
Classification accuracy, precision, recall, and F1-score measure the the agent's ability to correctly label dark patterns.
The explanation accuracy measures the agent's ability to provide supporting evidence and reasoning.
It is only measured when the agent has correctly identified a dark pattern.
The lower section reports the difference $\Delta$ between consecutive strategies, where the 95\% CI is computed using 1000 resamples of 50 ground truth workflows.  
$^*$~indicates a statistically significant change ($p < 0.05$).}
\label{tab:rq1:bootstrap_ablation}
\centering
\resizebox{\textwidth}{!}{%
\begin{tabular}{l cc cc cc cc cc}
\toprule
\textbf{Strategy} 
    & \multicolumn{2}{c}{\textbf{Class. Acc. (\%)}} 
    & \multicolumn{2}{c}{\textbf{Precision (\%)}} 
    & \multicolumn{2}{c}{\textbf{Recall (\%)}} 
    & \multicolumn{2}{c}{\textbf{F1 (\%)}} 
    & \multicolumn{2}{c}{\textbf{Expl. Acc. (\%)}} \\
\midrule

L1: Zero-shot
    & \multicolumn{2}{c}{70.8} 
    & \multicolumn{2}{c}{61.0} 
    & \multicolumn{2}{c}{60.1} 
    & \multicolumn{2}{c}{60.5} 
    & \multicolumn{2}{c}{78.1} \\

L2: Zero-shot + Role
    & \multicolumn{2}{c}{63.6} 
    & \multicolumn{2}{c}{49.9} 
    & \multicolumn{2}{c}{73.3} 
    & \multicolumn{2}{c}{59.4} 
    & \multicolumn{2}{c}{71.1} \\

L3: Few-shot + Role
    & \multicolumn{2}{c}{83.5} 
    & \multicolumn{2}{c}{79.9} 
    & \multicolumn{2}{c}{74.0} 
    & \multicolumn{2}{c}{76.9} 
    & \multicolumn{2}{c}{95.8} \\

L4: Few-shot + Role + CoT
    & \multicolumn{2}{c}{86.7} 
    & \multicolumn{2}{c}{88.0} 
    & \multicolumn{2}{c}{74.4} 
    & \multicolumn{2}{c}{80.7} 
    & \multicolumn{2}{c}{98.5} \\

\midrule
& \textbf{$\Delta$} & \textbf{95\% CI} 
& \textbf{$\Delta$} & \textbf{95\% CI} 
& \textbf{$\Delta$} & \textbf{95\% CI} 
& \textbf{$\Delta$} & \textbf{95\% CI} 
& \textbf{$\Delta$} & \textbf{95\% CI} \\
\midrule

$\Delta$ L1$\to$L2 (+Role)
    & $-7.1^*$      & [$-14.7$, $0.0$]
    & $-11.12^*$     & [$-21.9$, $-0.4$]
    & $+13.3^*$   & [$0.0$, $26.4$]
    & $-1.2$      & [$-11.1$, $8.6$]
    & $-6.9$      & [$-20.6$, $8.0$] \\

$\Delta$ L2$\to$L3 (+Few-shot)
    & $+19.9^*$   & [$13.5$, $26.7$]
    & $+30.0^*$   & [$19.2$, $41.0$]
    & $+0.7$      & [$-11.3$, $14.4$]
    & $+17.5^*$   & [$8.4$, $27.8$]
    & $+24.7^*$   & [$12.7$, $36.6$] \\

$\Delta$ L3$\to$L4 (+CoT)
    & $+3.2$      & [$-2.2$, $8.6$]
    & $+8.1$      & [$-1.7$, $19.2$]
    & $+0.4$      & [$-10.8$, $11.4$]
    & $+3.8$      & [$-5.1$, $11.7$]
    & $+2.7$      & [$-1.8$, $7.8$] \\

\bottomrule
\end{tabular}%
}
\end{table*}

\section{RQ1: Can LLM-Driven Agents Reliably Audit Dark Patterns?}
\label{sec:rq1}

In this section, we evaluate whether an LLM-driven browser agent can reliably identify dark patterns in right-to-access workflows under the standardized interaction protocol defined in \Cref{sec:design:groundtruth}.
We focus on two aspects of reliability using data obtained from our controlled prompt ablation study described in \Cref{sec:design:ablation}:
(1) the agent's ability to correctly classify the presence or absence of dark pattern categories relative to the human-annotated benchmark; and 
(2) the consistency and interpretability of the supporting explanations that it provides.
Finally, we present the results of our large-scale agent deployment to describe the prevalence and distribution of dark patterns in current right-to-access workflows.
The systematic constraints that limit this reliability are analyzed separately in Section~\ref{sec:rq2}.

\para{Workflow completion as a precondition for evaluating agent accuracy.}
Before evaluating the accuracy of classifications and explanations, we must assess whether the agent can meaningfully complete the right-to-access workflows using the standardized interaction protocol defined in \Cref{sec:design:groundtruth}.
Using the browser-use framework's internal {\tt is\_successful} flag, the agent reports a task completion rate of 95\% in Phase 2 and 89\% in Phase 3.
However, as described in \Cref{sec:design:ablation}, there are additional reasons for incomplete workflows that are not covered by the internal {\tt is\_successful} flag.
After applying the completion verification procedures described in \Cref{sec:design:ablation}, the actual workflow completion rate was determined to be 87\% in Phase 2 and 79\% in Phase 3.
This indicates that automated traversal of right-to-access workflows, using our defined interaction protocol, is feasible for a large majority of cases. 
We systematically analyze the reasons for failure in \Cref{sec:rq2}.
All subsequent accuracy metrics reported in this section are only over the successfully completed workflow executions --- \ie we evaluate agent accuracy conditioned on the successful completion of the workflow.

\subsection{Prompt configuration evaluation}
\label{sec:rq1:ablation}
We evaluate the four prompting configurations introduced in \Cref{sec:design:ablation} using the 100-sample ground truth dataset to identify the optimal configuration for dark pattern detection accuracy.
To assess whether performance differences between consecutive configurations are statistically meaningful, we conduct a bootstrap evaluation by repeatedly sampling 50 workflows with replacement from our ground truth dataset and computing the metric difference $\Delta$ between two consecutive strategies on each resample.
This sampling process is repeated 1000 times.
The resulting 95\% confidence intervals reflect the uncertainty of each metric difference across bootstrap samples.
\Cref{tab:rq1:bootstrap_ablation} reports the observed performance of each configuration (upper section) alongside the incremental effect $\Delta$ of each prompting component (lower section).

\para{Level 1: Zero-shot prompting.}
This configuration establishes the baseline.
It achieves 70.8\% classification accuracy (95\% CI: [66.0, 75.5]) and an F1-score of 60.5\% (95\% CI: [53.5, 67.5]).
Precision and recall are approximately balanced (61\% and 60.1\%, respectively), indicating that errors are somewhat evenly spread across false-positives and false-negatives.
The explanation accuracy is 78.1\% (95\% CI: [68.5, 86.3]), meaning that when correct assessments of dark patterns are made, the agent's rationale is reasonably aligned with the human-annotated evidence and mapping to the harm mechanism.
Importantly, the performance of this baseline suggests that pretrained models do encode some of the knowledge required to perform interaction-level dark pattern audits.
However, this alone is insufficient for high-fidelity audits.

\para{Level 2: Zero-shot prompting with a defined role.}
Adding regulatory role conditioning produces a marked shift in the precision-recall tradeoff over the baseline.
Recall experiences a statistically meaningful increase of 13.3 \textit{pp} (percentage points), while precision experiences a statistically meaningful decline of 11.1 \textit{pp}.
Despite these changes, the F1-score and explanation accuracy remain statistically indistinguishable from the baseline.
This result is counter to the common expectation that role framing improves overall task performance \cite{reynolds2021prompt}.
Instead, we observe increased sensitivity and false positives.
This suggests that role framing amplifies sensitivity of the reasonable expectation assessment at the cost of specificity.

\para{Level 3: Few-shot prompting with a defined role.}
Introducing example scenarios with explanations of how harm mechanisms observed in them map to dark pattern categories yields a substantial and statistically meaningful improvement to the agents performance.
The classification accuracy increases by 19.9 \textit{pp}, F1-score by 17.5 \textit{pp}, precision by 30 \textit{pp}, and explanation accuracy by 24.7 \textit{pp}.
All these improvements are statistically significant.
Only recall does not significantly improve relative to Level 2.
This indicates that the primary effect of few-shot prompting is not expanded detection coverage, but a sharp reduction in the false-positive rate.
Put differently, canonical examples derived from the ground truth data and prior literature appear to improve specificity while simultaneously maintaining the improved sensitivity from role framing.

\para{Level 4: Few-shot prompting with a defined role and Chain-of-Thought.}
Adding structured chain-of-thought reasoning yields the highest mean performance, with reasonable gains across the classification accuracy (3.2 \textit{pp}), explanation accuracy (2.7 \textit{pp}), precision (8.1 \textit{pp}), and F1-score (3.8 \textit{pp}).
In contrast, the recall remains stable with a marginal gain of only 0.4\%.
This result is contrary to common assumptions that chain-of-thought primarily improves recall on complex multi-step tasks \cite{wei2022chain}. 
However, our bootstrap analysis shows that these improvements over our Level 3 prompting strategy are not statistically significant.

\para{Takeaways.} Our analysis yields three important findings.
First, role framing alone increases recall but significantly worsens precision, yielding no net improvement in agent reliability.
Second, grounding the agent in a few in-context examples produces large and statistically significant improvements in classification accuracy, explanation accuracy, and F1-score (driven primarily through improvements in precision).
Finally, adding chain-of-thought reasoning yields the highest mean performance, primarily through improved precision and explanation quality, although the improvement over few-shot prompting is not statistically significant.
Based on these results, we select the Level 4 configuration for our agent.

\begin{table}[t]
\caption{{\it RQ1: Performance by dark pattern category under the selected Few-shot + Role + CoT configuration (\cf \Cref{sec:rq1:pattern}).}
}
\label{tab:rq1:pattern_level}
\centering
\resizebox{\columnwidth}{!}{%
\begin{tabular}{lcccccc}
\toprule
\textbf{Pattern} & \textbf{Class.} & \textbf{Prec.} & \textbf{Rec.} & \textbf{F1} & \textbf{Expl.} \\
                 & \bf Acc. (\%)  & \bf (\%) & \bf (\%) &\bf (\%)  & \bf Acc. (\%) \\
\midrule
Adding Steps             & 96.6 & 95.2 & 90.9 & 93.0 & 100.0 \\
Conflicting Info.  & 88.6 & 95.7 & 71.0 & 81.5 & 100.0 \\
Creating Barriers        & 86.7 & 91.3 & 85.7 & 88.4 & 97.6 \\
Feedforward Ambiguity    & 86.8 & 94.1 & 76.2 & 84.2 & 100.0 \\
Hidden Info.       & 69.9 & 85.0 & 40.5 & 54.8 & 94.1 \\
Info. Without Context     & 87.6 & 93.3 & 75.7 & 83.6 & 100.0 \\
Privacy Mazes            & 81.8 & 61.5 & 72.7 & 66.7 & 93.8 \\
Visual Prominence        & 95.7 & 84.0 & 100.0 & 91.3 & 100.0 \\

\bottomrule
\vspace{-7mm}
\end{tabular}}
\end{table}

\subsection{Pattern-specific performance evaluation}
\label{sec:rq1:pattern}

We also examine detection performance at the level of individual dark pattern categories under the selected prompting configuration (Level 4: Few-shot + Role + CoT).
\Cref{tab:rq1:pattern_level} reports the per-category classification accuracy, precision, recall, F1-score, and explanation accuracy using the 100-sample ground truth dataset as the reference.
At a high-level, we find that detection performance varies systematically across dark pattern categories, and this variation aligns closely with how interaction-dependent each pattern is.
More specifically, we make four key observations.

\para{Patterns that manifest through immediately observable interface elements are detected with high reliability.}
Specifically, {\em Adding steps} achieves a 96.6\% classification accuracy and 93\% F1-score, with both precision (95.2\%) and recall (90.9\%) remaining high.
Similarly high performance is also seen in the {\em Visual prominence} category.
In both these categories, the underlying pattern that leads to their occurrence is structurally localized --- \eg additional required fields, redundant steps, or visually emphasized controls are directly exposed within a single interaction state or screen.
This suggests that when the signal is explicit and bounded within a page-level context, the agent is able to consistently detect it.

\para{Patterns involving textual contradictions exhibit high precision and moderate recall.}
Patterns such as {\em Conflicting information}, {\em Feedforward ambiguity}, and {\em Information without context} all achieve remarkably high precision (93\%+) but significantly lower recall (71-76\%).
These categories require reconciling textual information that may occur across sections and pages.
While such contradiction-related patterns are usually correct when identified, their detection may often depend on the agent's ability to maintain context and compare content across steps within a workflow.

\para{The most challenging patterns are those that are inherently interaction-dependent.}
The {\em Privacy mazes} and {\em Hidden information} patterns exhibit the lowest performance with F1-scores of 66.7\% and 54.8\%, respectively.
Notably, {\em Hidden information} exhibits high precision (80.5\%) but very low recall (40.5\%), indicating that the agent is conservative in assigning the label and frequently fails to detect it.
Interestingly, both categories can be considered inherently `interaction-dependent' --- \ie they often manifest when users traverse multiple pages, encounter fragmented instructions, or discover missing disclosures over the course of navigating the workflow.
One hypothesis for this lower performance is that the agent's primary difficulty is not in recognizing these patterns, but in aggregating distributed signals across multi-step workflows.

\para{Explanation accuracy is uniformly high across all dark pattern categories.}
Across all categories, explanation accuracy exceeds 93\%. 
This indicates that when the agent does identify an interaction-level violation, its supporting rationale consistently references observable interface elements and aligns with the predefined harm mechanism mapping. 
The principal limitation facing the agent is therefore not justification quality, but incomplete detection of patterns that emerge only through workflow navigation and maintenance of context.

\para{Takeaways.}
Overall, the results suggest that the agent's strongest performance occurs when dark patterns are structurally localized and directly observable. 
Detection becomes more difficult when patterns depend on multi-step navigation, contextual comparison, or inference from omissions. 
Patterns whose signals are distributed across multiple workflow states present the greatest challenge for automated auditing under a fixed interaction protocol.

\begin{table}[t]
\caption{
{\it RQ1: Dark pattern prevalence estimates (\cf \Cref{sec:rq1:deployment}).}
Ground truth prevalence is computed from the 100-sample ground truth dataset.
Deployment prevalence reflects observations across the remaining 356 brokers.
}
\label{tab:rq1:deployment}
\centering
\resizebox{\columnwidth}{!}{%
\begin{tabular}{l cc cc}
\toprule
\textbf{Pattern} 
& \multicolumn{2}{c}{\textbf{Ground Truth}} 
& \multicolumn{2}{c}{\textbf{Deployment}} \\
\cmidrule(lr){2-3} \cmidrule(lr){4-5}
& \textbf{Prev. (\%)} & \textbf{95\% CI} 
& \textbf{Prev. (\%)} & \textbf{95\% CI} \\
\midrule
Adding Steps            & 24.1 & [14.9, 33.3] & 15.2 & [11.1, 19.4] \\
Conflicting Info. & 26.1 & [17.0, 35.3] & 20.1 & [15.4, 24.9] \\
Creating Barriers       & 55.6 & [44.7, 66.4] & 48.6 & [42.9, 54.4] \\
Feedforward Amb.   & 37.8 & [27.8, 47.8] & 25.2 & [20.0, 30.4] \\
Hidden Info.      & 21.1 & [12.7, 29.5] & 21.1 & [16.2, 26.0] \\
Info. w/o Context & 33.7 & [23.9, 43.5] & 34.5 & [28.9, 40.1] \\
Privacy Mazes           & 26.8 & [17.2, 36.4] & 20.1 & [15.5, 24.6] \\
Visual Prominence       & 27.8 & [18.5, 37.0] & 20.0 & [15.2, 24.8] \\
\bottomrule
\end{tabular}%
}
\end{table}

\subsection{Observations from large-scale deployment}
\label{sec:rq1:deployment}

We deploy the Level 4 prompting configuration (Few-shot + Role + CoT) across the remaining 356 data brokers to obtain an estimate of the prevalence of dark patterns in current right-to-access workflows.
Prevalence is computed only for workflows that satisfy the verification criteria described in \Cref{sec:design:ablation}.
Data brokers for which the workflows could not be completed are excluded from the denominator.
\Cref{tab:rq1:deployment} summarizes our results.
The prevalence rates and intervals reported in the `Ground Truth' columns are computed from the 100-sample ground truth dataset, while the rates and intervals reported in the `Deployment' columns are computed from our large-scale deployment.

\para{Structural barriers are the most prevalent dark pattern.}
We found that {\em Creating barriers} was identified by our agent in roughly one-half of all completed workflows across both datasets.
The most common manifestations included mandating completion of non-essential fields prior to request submission (\eg performing biometric verification and providing government IDs) or mandating a shift of modality (\eg requiring the download of an external mobile app to complete the submission request).
Based on our evaluation with reference to the ground truth dataset, our agent exhibits strong performance (F1-score: 88.4\%) in this category, suggesting that the observed prevalence of these artificial barriers is unlikely to be a poor estimate resulting from significant over- or under-counting.

\para{Ambiguity- and fragmentation-based patterns are widespread.}
Several categories occur in roughly one-quarter to one-third of our completed workflows.
These include {\em Feedforward ambiguity}, {\em Information without context}, {\em Conflicting information}, and {\em Privacy mazes}.
These patterns commonly manifest as instruction-outcome mismatches (\eg links advertised as access requests that lead only to opt-out workflows), contextual misplacement of information (\eg placing links to submission portals in the ``Contact us'' section of a privacy policy rather than in the ``Data rights'' section), and cross-page fragmentation of submission instructions.
In our evaluation, our agent generally exhibited high precision but lower recall across these categories, suggesting that our reported prevalence rates for these categories are likely under-estimating their true prevalence.

\para{Takeaways.}
Overall, we find that dark patterns are widespread in right-to-access workflows.
Structural barriers emerge as the dominant pattern, while ambiguity- and fragmentation-related patterns are also widespread. 
These prevalence estimates are conditioned on verified workflow completion and on our agent's pattern detection capabilities, and therefore represent estimates rather than verified prevalence rates.
Importantly, the interaction-dependent nature of several high-prevalence categories implies that meaningful auditing requires traversal and contextual aggregation rather than static inspection of isolated interface elements.

\section{RQ2: What Constrains LLM-Driven Agents?}
\label{sec:rq2}

Our results in \Cref{sec:rq1} demonstrate that LLM-driven agents can reliably identify dark patterns in the right-to-access workflows.
However, to understand the practical limits of autonomous auditing, we must examine when the agent fails.
These failures may arise due to incomplete workflows (\Cref{sec:rq2:execution}). When the workflow is completed, interaction-level auditing may still be limited by unobserved interface signals (\Cref{sec:rq2:observation}) or reasoning errors (\Cref{sec:rq2:reasoning}), which help explain the lower detection accuracy for certain dark patterns.
In this section, we analyze these constraints systematically, distinguishing execution limits from observation and reasoning failures.

\subsection{Workflow execution limits}
\label{sec:rq2:execution}
A fundamental constraint on autonomous auditing is the ability to meaningfully complete multi-step interaction workflows.
Our agent is able to complete 81\% of all workflows using the defined interaction protocol (87\% in Phase 1 and 79\% in Phase 2). 
The remaining workflows cannot be completed and therefore cannot be audited for dark patterns autonomously.
They represent the coverage limits of the agent as an auditing instrument.
Examining the causes for these failures using the failure categories described in \Cref{sec:design:ablation}, we note the following  observations and highlight their implications for agent design. 

\para{Infrastructure and security barriers account for most workflow execution failures.}
Among the incomplete workflows, automation instability accounts for 26.7\% of all failures and security barriers (\eg CAPTCHAs) account for 25.8\%.
Together, these account for 52.5\% of all failed workflow executions.
These failures arise from environmental constraints external to the model's reasoning process.
Consequently, even a perfectly calibrated dark pattern classifier would remain unable to evaluate these workflows without modifications to the underlying system infrastructure and agent-website interaction mechanics.
The high prevalence of infrastructure and security-related failures indicates that agent coverage is constrained by the stability of the browser instrumentation, network infrastructure, and the presence of anti-both technologies on data broker websites.
This constraint is consistent with prior agent research demonstrating that web automation is sensitive to factors such as asynchronous content loading  and security mechanisms \cite{zhou2023webarena}.
Improving coverage in this setting likely requires the development of session recovery mechanisms that allow agents to detect failures, replan, and resume from intermediate checkpoints \cite{shinn2023reflexion}.
Covert crawling strategies (\eg disguising user agents, using residential proxies, \etc) could also improve workflow completion rates and representativeness of data \cite{ahmad2020apophanies}.
However, such approaches raise ethical and methodological concerns.
Circumventing anti-bot mechanisms risks violating deployment norms and platform safeguards, thereby imposing a principled limit on achievable coverage.

\para{Dynamic and multi-step interactions increase agent fragility.}
Interaction failures account for 24.4\% of all incomplete workflows.
In these cases, the agent is able to locate a submission pathway but unable to fully expose required fields or conditional steps in multi-stage forms.
Navigation failures account for an additional 10.1\% of failures, reflecting cases where the agent is unable to operationalize the defined interaction protocol.
Content format limitations contribute to 7.8\% of failures, typically when relevant instructions are embedded in formats such as PDFs which our agent is unable to interact with.
Together, these reflect a substantial portion of failures (42.4\%) which arise primarily due to the interaction-level complexity of right-to-access workflows.
Specifically, because many right-to-access workflows reveal additional fields or instructions conditionally, reliable auditing requires parallel exploration of alternative form states and navigation branches rather than following a single linear interaction path.
Although agents operating in the ReAct-style paradigm \cite{yao2022react}, such as ours, interleave reasoning and action, the reliable traversal of conditional and branching workflows that are encountered in right-to-access workflows requires the agent's state management to allow for explicit branch enumeration and exploration capabilities.
We expect that extending agent capabilities to enumerate and explore these alternate form states will directly reduce the prevalence of such failures.

\para{Internal agent instability represents a small fraction (5.1\%) of failed workflows.}
These failures arise from inconsistencies in the agent's internal planning state rather than external barriers.
While comparatively infrequent, these errors suggest that there are still improvements to be made to agents' internal state management.
Specifically, techniques such as schema enforcement and intermediate state verification, with the ability to resume from prior checkpoint states would likely reduce these types of failures.

\para{Takeaways.}
Taken together with our previous results, it appears that improvements to interaction-level workflow traversal and internal state management enhancements will likely have a larger impact on agent usability, in comparison to improving dark pattern classification capabilities.
This is because once relevant signals are surfaced, the agent exhibits strong classification and explanation performance. 
The primary bottleneck lies, instead, in reliably exposing these signals in complex web environments.

\subsection{Agent observation limits}
\label{sec:rq2:observation}
Even when workflows are successfully executed, reliable auditing requires agents to surface and perceive relevant interface signals. 
Our results in ~\Cref{sec:rq1:pattern} show that our agent suffers low recall for several patterns.
This is a potential indication that failures may occur not because the agent lacks knowledge of dark patterns, but because these signals are not consistently exposed during interaction.
Here, we highlight these failures and their implications for agent design.

\para{Visually subtle patterns not always reliably perceived by the agent.}
This observation is most pronounced in the {\em Hidden information} pattern, which exhibits high precision (85\%) and low recall (40.5\%).
In many cases, this pattern manifests as visually disguised hyperlinks, non-actionable references, or information hidden within expandable elements.
Although screenshots are supplied to the agent's underlying LLM, reliable detection of these artifacts requires that:
(1) the relevant elements have to be surfaced (\eg expandable sections must be triggered), and 
(2) the agent must integrate these visual cues with the structured DOM information.
Prior work on multi-modal web agents has shown that vision grounding improves interface reasoning but does not always result in consistent alignment between this reasoning and the rendered content \cite{zhou2023webarena}.
From an agent design perspective, this suggests that multi-modal reasoning is necessary but insufficient.
Reliable auditing may require explicit `expand-all' strategies for collapsible content, and post-action DOM–vision consistency checks to ensure that relevant states have been exposed before classification.

\para{Fragmented signals challenge agent context window limitations.}
Patterns such as {\em Privacy mazes} and {\em Conflicting information} require comparing instructions from multiple navigation steps.
These patterns depend on aggregating signals that are individually unimportant, but collectively inconsistent.
We hypothesize that these errors manifest due to context window limitations in the underlying LLM, where the earlier observations may be compressed or summarized --- a known challenge in agents that rely on the ReAct-style paradigm \cite{yao2022react}.
Notably, our results match findings from prior work which indicate that this limitation can cause performance degradation in web agents performing long interactions \cite{deng2023mind2web, zhou2023webarena}.
This hypothesis is further supported by the agent's strong performance in detecting dark patterns which typically involve directly observable artifacts that appear within a single navigation step (\eg {\em Adding steps} or {\em Creating barriers}), suggesting that the reliable exposure and aggregation of fragmented signals results in this limitation.
Design implications here center on structured memory management. 
Specifically, agents may benefit from persistent state representations that identify and store key disclosures explicitly, allowing for the detection of patterns that require long interaction sequences and the piecing together of fragmented signals.

\para{Takeaways.}
Taken together, these findings indicate that observation limits arise from the interaction-heavy nature of modern web interfaces.
Neither multi-modal input nor additional prompt configuration can eliminate blind spots if pages, elements, and patterns are not fully visible or if key information is removed from or compressed in the underlying model's context window.
Thus, improvements in autonomous auditing will likely require stronger integration between reasoning and multi-modal inputs, as well as strategic context retention.

\subsection{Agent reasoning limits} \label{sec:rq2:reasoning}

Even when signals are successfully exposed, reliable auditing requires the agent to correctly interpret and reason about them. 
Our results in \Cref{sec:rq1:pattern} show several patterns with lower precision, potentially indicating failures due to an incorrect or overly simplified understanding of the dark pattern.
Here, we highlight these failures and their implications for agent design.

\para{Some dark patterns require interpretive judgments, which challenge agent performance.}
Several dark pattern subtypes depend on assessments of whether a design choice is reasonable or excessive.
For example, identity verification requirements may be appropriate for fraud prevention, yet demanding government-issued identification or biometric verification may be seen as imposing disproportionate burden.
In such cases, the agent must evaluate proportionality rather than merely identifying the presence of a feature.
Errors here do not stem from the failure to observe specific elements or features, but from the difficulty in determining the (ambiguous) boundaries between protection and obstruction.
From an agent design perspective, similar to prior research \cite{yao2022react, bai2022constitutional}, we hypothesize that the agent could benefit from separating factual observations from interpretive evaluations (of proportionality), potentially allowing for more stable judgments.
In addition, the agent could be designed to flag borderline cases as uncertain rather than forcing a binary classification.
Importantly, such interpretive judgments are not unique to automated systems.
Research on dark pattern identification routinely involves normative assessments of user expectations and reasonable design practice \cite{gray2018dark}.
The challenge, therefore, is not eliminating normative judgment, but structuring it in a way that is transparent, reproducible, and systematically applied within the agent’s reasoning process.

\para{Reasoning instability also arises when multiple weak signals must be aggregated across steps.}
As previously discussed, some dark pattern categories emerge from the cumulative effect of multiple fragmented signals.
In these cases, the agent must combine evidence across navigation steps and determine whether the aggregated structure meets the threshold for a dark pattern.
While poor recall in such detection can be explained by information loss due to context window limitations, lower precision suggests that errors occur not because the signals are absent, but because the agent's reasoning mechanism is unstable.
From a design perspective, these findings suggest that reasoning over interaction-level workflows would benefit from explicit state representation rather than purely narrative aggregation. 
Similar to above, this may be achieved through separation of information retrieval and judgment phases.
Put differently, instead of incrementally forming judgments as the workflow unfolds, the agent could maintain a structured record of key observations (\eg detected submission channels, required identifiers, and stated verification requirements) throughout traversal, and use these to make interpretive judgments.

\para{Takeaways.}
Taken together, these findings indicate that reasoning limits in interaction-level dark pattern auditing arise from two distinct sources: interpretive ambiguity and cumulative evidence aggregation. 
Some categories require proportionality judgments that do not admit sharp boundaries, while others require combining distributed signals across multiple navigation steps. 
In both cases, errors arise not from missing signals, but from the difficulty of applying consistent judgment in situations where the structure of the workflow and the normative standards for evaluation are inherently ambiguous.
Improving reliability will therefore require more explicit separation between factual observation and evaluative reasoning, structured aggregation of multi-step evidence, and principled handling of borderline cases.
\section{Related Work}  \label{sec:related}

\para{Dark patterns in privacy choices.}
Within the privacy domain, dark patterns manifest as barriers that cause users to unintentionally disclose personal information~\cite{bosch2016tales} or impede consent withdrawal and subscription cancellation~\cite{sheil2024staying}.
Empirical audits of cookie banners reveal widespread use of nudging and asymmetric choice architectures that bias users toward consent \cite{krisam2021dark, nouwens2020dark}, while studies of opt-out mechanisms document friction and asymmetry that undermine meaningful choice~\cite{o2021clear}.
Beyond interface visibility, Tran \etal ~\cite{tran2025dark} document the full opt-out submission and follow-up process, showing that a substantial portion of requests fail, remain ambiguous, or require excessive additional steps.
Parallel findings emerge in studies of the Right of Access (ROA). 
Pöhn \etal ~\cite{pohn2023needle} conduct a systematic analysis and report widespread obstructive dark patterns in GDPR-based access workflows.
Lobel \etal ~\cite{lobel2024access} extend this analysis at larger scale and find that more than 68\% of right-to-access workflows embed at least one dark pattern, with obstruction-oriented mechanisms being the most prevalent.
Our study extends this line of research by systematically auditing dark patterns in how California-registered data brokers operationalize the right-to-access process at scale.

\para{Dark pattern detection and LLM agents.}
Prior detection approaches have evolved from HTML crawling and 
clustering~\cite{mathur2019dark, nazarov2022clustering} to computer vision and NLP-based screenshot analysis~\cite{chen2023unveiling, mansur2023aidui}, to multimodal LLMs that generate structured explanations from static UI artifacts~\cite{shi202550, kocyigit2025deceptilensc}.
The shortage is they operate on static representations and cannot capture patterns that emerge through multi-step interaction.
LLM-driven browser agents have demonstrated the ability to execute realistic end-to-end workflows across diverse web environments \cite{zhou2023webarena, deng2023mind2web, he2024webvoyager, browseruse2025, multion2024, skyvern2025}.
However, despite these advances, LLM agents remain vulnerable along multiple dimensions. 
Recent work systematically analyzes agent failures to understand when and why agents break, and how errors propagate across steps.
These studies examine erroneous planning behaviors.
erroneous planning \cite{ji2024testingunderstandingerroneousplanning}, tool-use errors~\cite{ning2024definingdetectingdefectslarge}, hallucination ~\cite{lin2025llm}. 
Recent work shows that these failures rarely remain isolated, instead 
cascading across planning, memory, and action modules in ways that 
compound task failure~\cite{zhu2025llmagentsfaillearn}.
Our work builds on this literature by empirically evaluating both the capabilities and boundary conditions of LLM-driven agents in the context of regulatory dark pattern auditing.

\section{Discussion and Conclusions} \label{sec:discussion}
\para{Implications for scalable regulatory auditing.} 
Our evaluation demonstrates that LLM-driven agents can feasibly support scalable dark pattern auditing. 
The agent successfully completes 81\% of workflows and, under the best-performing configuration, achieves strong classification and explanation accuracy at an average cost of approximately \$0.52 ($\approx$287k tokens) per data broker.
At the same time, reliability remains bounded along three dimensions. 
First, execution coverage is constrained by infrastructure and security-related failures (e.g., automation timeouts or verification mechanisms), and interaction-level failures with dynamic web elements.
Second, detection reliability depends critically on carefully constructed domain-specific few-shot grounding and reasoning scaffolds.
Third, robust end-to-end auditing right-to-access workflows require agents to maintain structured memory for aggregating distributed information and explicit DOM–vision alignment to reliably detect cross-step contextual and visually subtle dark patterns. 
Taken together, LLM-driven agents are better deployed as scalable compliance triage tools. 
They can autonomously traverse workflows and flag dark patterns with structured evidence, while human reviewers responsible for adjudicating borderline cases.
More broadly, our results demonstrate a viable pathway for large-scale dark pattern auditing in the privacy domain. With appropriate task-specific grounding, similar architectures may be extended to other regulatory rights workflows, including deletion and opt-out flows.\\

\para{Conclusion.} We evaluated the feasibility and limitations of LLM-driven agents for interaction-level dark pattern auditing, using CCPA right-to-access portals as a testbed.
We design and deploy an agent built on the browser-use framework with GPT-5, demonstrating that it can traverse complex multi-step workflows, produce evidence-backed classifications, and achieve strong performance when supported by structured few-shot examples and chain-of-thought reasoning.
Across the full deployment, we find that structural barriers are the most prevalent dark pattern, present in roughly half of all completed workflows, while ambiguity- and fragmentation-based patterns appear in one-quarter to one-third.
At the same time, we identify systematic boundary conditions arising from workflow execution failures, driven primarily by security barriers, automation instability, and incomplete interaction with dynamic multi-step interfaces.
Observation and reasoning limits further constrain performance on patterns that require aggregating distributed signals or making proportionality judgments. 
These findings collectively characterize the promise and current boundaries of LLM-driven dark pattern auditing, and establish a principled foundation for evaluating and improving agentic auditing systems in regulatory contexts.

\clearpage
\bibliographystyle{ACM-Reference-Format}
\bibliography{reference}

\appendix
\section{Appendix} \label{sec:appendix}

\subsection{Dark Pattern Definitions and Harm Mechanisms} \label{sec:appendix:harm_mapping}
We provide full details of the eight high frequency dark pattern definitions, along with the associated harm mechanisms observed in the 100 annotated data broker workflows, as referenced in \Cref{sec:design:groundtruth}.
\para{Adding Steps:} Subverts the user’s expectation that a task will take as few steps as technologically necessary, instead introducing additional required interactions.
\begin{itemize}
    \item \textit{Identifier fragmentation.} Users are required or encouraged to submit separate requests for different identifiers (e.g., email address, IP address, cookie ID, or mobile advertising identifier). Because the interface accepts only one identifier per submission, users with multiple identifiers must repeat the process multiple times. 
    
    \item \textit{Request-type fragmentation.} A single legal access right is split into multiple sub-request categories (e.g., “categories of personal information” vs. “specific pieces of personal information”). To obtain a complete disclosure, users must either submit multiple requests or correctly interpret and select several overlapping options if companies allow multiple choice in selection. 
\end{itemize}

\para{Conflicting Information:} Includes two or more sources of information that conflict with each other, creating uncertainty about the consequences of the user’s actions. Recurring subtypes in ROA workflows:
\begin{itemize} 

    \item \textit{Submission path conflict.} Different pages present incompatible instructions about valid submission channels. For example, one page lists email as an official CCPA method while another claims only a webform is accepted. Such contradictions force users to guess which instruction is authoritative.
    
    \item \textit{Submission scope contradiction.} A policy claims that a linked pathway supports multiple CCPA rights, but the destination interface supports only a subset (e.g., opt-out only). Users are misled into believing they can exercise access rights through a path that does not actually provide them.
    
    \item \textit{Form labeling contradiction.} Similar, some brokers display a link advertised as a general or access-right request leads to a page prominently labeled as “Opt-Out” or “Do Not Sell”.
\end{itemize}

\para{Creating Barriers:} Subverts the user’s expectation that relevant tasks will be supported by the interface, instead preventing or unnecessarily complicating the task to disincentivize user action. We observe several recurring subtypes in ROA workflows:
\begin{itemize}
    \item \textit{Role classification barriers.} Users are required to select their relationship to the company from a rigid predefined list (e.g., “customer,” “participant,” “professional”). Consumers who hold multiple roles or cannot identify their category face uncertainty and risk misrouting their request.
    
    \item \textit{Submission channel restriction.} The company provides only a single submission method (e.g., webform only or email only), while CCPA require to provide at least two submission methods [1798.130.(a).(1)]\cite{CCPA}.

    \item \textit{Excessive identity verification.} Requests require disproportionately sensitive or difficult to obtain materials such as government-issued IDs, SSNs, biometric verification, or physical mail authentication. These requirements exceed what is necessary for normal verification.

    \item \textit{Technical identifier burden.} Users are required to provide device-level identifiers such as cookie IDs or mobile advertising identifiers, which are difficult for ordinary consumers to locate and interpret.

    \item \textit{Non-essential required fields.} Forms mandate submission of information unrelated to verification or request fulfillment, such as submit supporting documents or account numbers.

    \item \textit{Install external App.} Users are instructed to download third-party applications or tools to retrieve required identifiers like MAID, introducing technical complexity and additional privacy concerns.

    \item \textit{Self-identification burden.} Some people-search sites require consumers to locate and submit their own records (e.g., profile URLs) when filing a request. However, some search results are ambiguous, outdated, increasing effort and discouraging completion of the access request.
    
\end{itemize}

\para{Feedforward Ambiguity:} Subverts the user’s expectation that their choice will result in a predictable outcome, instead creating a discrepancy between provided information and the resulting action. Subtypes in ROA workflows are:
\begin{itemize}

    \item \textit{Instruction–outcome mismatch.} Links or interface instructions promise a general privacy request pathway, but the destination interface supports only a limited subset of rights or redirects users into an unrelated workflow (e.g., account creation). The resulting process violates the user’s expectation of a direct and predictable access request.

    \item \textit{Ambiguous rights terminology.} Policies and webforms use non-standard or vague labels (e.g., “Info Request” or “Data Processing”) that do not clearly map to legally defined access rights. Without explanation, users may hard to predict what selecting an option will actually trigger.

    \item \textit{Coupled outcomes.} Submitting an access request, but the destination shows that it will automatically trigger additional actions (such as opt-out).
\end{itemize}

\para{Hidden Information:} Subverts the user’s expectation that relevant information will be accessible and visible, instead disguising it or framing it as irrelevant.

\begin{itemize}
    \item \textit{Visually disguised affordances.} Clickable elements are present but styled to resemble ordinary text, lacking standard hyperlink affordances such as color contrast or emphasis. Therefore, users scanning the page cannot easily recognize that an actionable control exists.

    \item \textit{Non-actionable references.} We observed many brokers' policies reference information using plain text without providing direct navigation like hyperlinked. Users must manually search or scroll through lengthy documents to locate the resource.

    \item \textit{Interaction-gated disclosure.} Rights instructions are hidden behind expandable sections, “read more” links, or hover-only elements.

    \item \textit{Selective omission.} A section framed as comprehensive and dedicated to privacy rights or how to exercise them. However, it omits some information, such as not providing full submission methods, where that information exists elsewhere on the site without cross-references. Users who reasonably assume the section is complete may stop searching and miss critical information needed to exercise their rights.
\end{itemize}

\para{Information Without Context:} Alters relevant information or user controls in ways that limit discoverability, making it unlikely that users will find the information or action they seek. 

\begin{itemize}
    \item \textit{Contextual misplacement of execution guidance.} Access submission instructions appear in unrelated sections (e.g., "Contact Us" or "opt-out" section) and are not referenced within the section users reasonably expect to be. As a result, consumers encounter incomplete guidance unless they search beyond the expected context.
    
    \item \textit{Contextual misleading form framing.} Access-capable forms are labeled as opt-out or marketing controls. Even when access is technically supported, the framing signals irrelevance and discourages users from using the form for access requests.

    \item \textit{Vague or non-standard request labels.} Access request options use terminology (e.g., “Data Processing” “Info Request”) that does not clearly map to standard legal term.

    \item \textit{Section label mismatch.} Policies direct users to a named section for CCPA information, but the actual content appears under a differently labeled section without explanation.

    \item \textit{Misdirect pathways.} Some interfaces provide a privacy mechanism but fails to provide the limiting context (e.g., the webform is only for B2B clients or only for other different legal regime). By presenting a narrow tool under a general context, leading users believe they have reached the correct pathway but cannot complete the intended action.
\end{itemize}

\para{Privacy Mazes:} Requires users to navigate through multiple screens or fragmented pathways to accomplish a task, without providing a clear or comprehensive overview. 

\begin{itemize}
    \item \textit{Excessive navigational depth.} Users must follow a chain of intermediate pages before reaching the actual submission interface.
    
    \item \textit{Within-page fragmentation.} Access instructions are split across separate sections of the same page, and no single section presents a complete set of submission methods.

    \item \textit{Cross-page fragmentation.} Submission guidance is distributed across different pages, each containing partial instructions. No page provides an authoritative summary, forcing users to traverse multiple documents to assemble a complete understanding.
\end{itemize}

\para{Visual Prominence:} Places an element relevant to user goals in visual competition with a more distracting and prominent element, thereby diverting attention from the user’s intended task.
\begin{itemize}
    \item \textit{Competing call-to-action dominance.} Prominent buttons or banners (e.g., account creation prompts) visually outweigh the access request pathway, which appears as a small or low-contrast link. The imbalance draws attention away from the user’s primary task.

    \item \textit{Persistent overlay interference.} Floating widgets or fixed interface elements remain on screen and overlap privacy controls or footer links, obstructing interaction and diverting users attention.

\end{itemize}

\subsection{Prompt}
This section presents the complete prompt configuration used across the four prompting strategies (Level 1 to Level 4), as referenced in \Cref{sec:design:ablation}. This prompt is organized into functional sections, each governing a distinct aspect of the auditing task. In the figure, each section header is annotated with a concise description (shown as a colored label on the right) summarizing its functional role for clarity. For example, [GOAL] specifies the agent objectives, and [INTERACTION PLAN] prescribes the step-by-step website navigation and form interaction sequence. The four prompting strategies are also color-coded for clarity.

\label{sec:appendix:prompt}

\begin{figure*}[t]
  \centering
  \resizebox{\textwidth}{!}{
    \usetikzlibrary{positioning}

\definecolor{orangelight}{RGB}{255, 220, 170}
\definecolor{orangeborder}{RGB}{200, 140, 50}
\definecolor{headerbg}{RGB}{242, 242, 242}
\definecolor{greenlabel}{RGB}{180, 230, 180}
\definecolor{yellowlabel}{RGB}{255, 230, 180}
\definecolor{pinklabel}{RGB}{255, 200, 200}
\definecolor{innerboxbg}{RGB}{252, 248, 230}
\definecolor{rowtwocolor}{RGB}{255, 240, 200}
\definecolor{rowthreecolor}{RGB}{255, 220, 170}
\definecolor{innerborder}{RGB}{180, 160, 100}
\definecolor{bluelabel}{RGB}{200, 215, 240}

\definecolor{colorL1}{HTML}{FFE6CC}
\definecolor{colorL2}{HTML}{FAD7AC}
\definecolor{colorL3}{HTML}{FFD9A0}
\definecolor{colorL4}{HTML}{FFAD5C}

\begin{tikzpicture}[
  every node/.style={font=\small},
  roundtag/.style={
    rounded corners=4pt,
    inner xsep=6pt, inner ysep=3pt, font=\small\bfseries
  },
  smalltag/.style={
    rounded corners=3pt,
    inner xsep=4pt, inner ysep=2pt, font=\footnotesize\bfseries
  },
  labeltag/.style={
    rounded corners=3pt,
    inner xsep=5pt, inner ysep=3pt, font=\footnotesize
  }
]

\def\W{18.5}       
\def\TW{17.7cm}  

\draw[draw=gray!70, rounded corners=4pt, fill=white]
  (0,0) rectangle (\W, -17.8);

\draw[draw=gray!60, fill=headerbg]
  (0,0) rectangle (\W,-0.9);
\node[anchor=west, font=\small\bfseries] at (0.3,-0.45) {Prompt Strategies:};

\node[roundtag, draw=colorL1!80!black, fill=colorL1] (l1) at (4.3, -0.45) {L1: Zero-shot};
\node[roundtag, draw=colorL2!80!black, fill=colorL2] (l2) at (7.5, -0.45) {L2: Zero-shot + Role};
\node[roundtag, draw=colorL3!80!black, fill=colorL3] (l3) at (11.1,-0.45) {L3: Few-shot + Role};
\node[roundtag, draw=colorL4!80!black, fill=colorL4] (l4) at (15.1,-0.45) {L4: Few-shot + Role + CoT};

\draw[gray!60] (0,-0.9) -- (\W,-0.9);

\node[anchor=west, font=\small\bfseries] at (0.3,-1.2) {[ROLE]};
\node[anchor=west, font=\small\itshape] at (1.55,-1.2) {(not present in L1)};
\node[smalltag, draw=colorL2!80!black, fill=colorL2] at (4.3,-1.2) {L2};
\node[smalltag, draw=colorL3!80!black, fill=colorL3] at (5.1,-1.2) {L3};
\node[smalltag, draw=colorL4!80!black, fill=colorL4] at (5.9,-1.2) {L4};

\draw[gray!40] (0,-1.55) -- (\W,-1.55);

\node[anchor=north west, text width=\TW, font=\small] at (0.3,-1.65) {%
You are an expert auditor working for a regulatory agency, tasked with conducting a thorough investigation to identify all dark patterns on: \texttt{\$url}.
Your responsibility is to detect, document, and justify every potential dark pattern during the process of exercising and evaluating CCPA Right of Access.
};

\draw[gray!60] (0,-2.9) -- (\W,-2.9);

\draw[draw=gray!60, fill=headerbg]
  (0,-2.9) rectangle (\W,-3.55);
\node[anchor=west, font=\small\bfseries] at (0.3,-3.22) {[GOAL]};
\node[labeltag, draw=gray!50, fill=bluelabel, anchor=east]
  at (18.3,-3.22) {Specify objectives must accomplish};
\draw[gray!40] (0,-3.55) -- (\W,-3.55);

\node[anchor=north west, text width=\TW, font=\small] at (0.3,-3.8) {%
Your goal is to follow the [INTERACTION PLAN] to evaluate the process of exercising a CCPA Right of Access request in order to:\\[2pt]
(1) Evaluate clarity, completeness, and accuracy of CCPA Right of Access as described under [EVALUATION INSTRUCTIONS].\\[2pt]
(2) Comprehensively detect and classify all applicable dark patterns using [TAXONOMY], by applying [DECISION RULES]. Be exhaustive. Do not conclude the analysis until every observed aspect of the webpage or navigations or user actions have been compared against the entire taxonomy.\\[2pt]
(3) Collect and record verifiable, structured evidence that allows a human annotator to confirm your reasoning, as specified in [EVIDENCE REQUIREMENTS].
};

\draw[gray!60] (0,-6.85) -- (\W,-6.85);

\draw[draw=gray!60, fill=headerbg]
  (0,-6.85) rectangle (\W,-7.5);
\node[anchor=west, font=\small\bfseries] at (0.3,-7.175) {[INTERACTION PLAN]};
\node[labeltag, draw=gray!50, fill=bluelabel, anchor=east]
  at (18.3,-7.175) {Prescribe the step-by-step website navigation and form interaction sequence (See \S3.1))};
\draw[gray!40] (0,-7.5) -- (\W,-7.5);

\node[anchor=north west, text width=\TW, font=\small] at (0.3,-8.0) {%
Follow the interaction plan listed below while attempting to exercise CCPA Right of Access on the provided website.\\[2pt]
(Step 1) Find all links on the provided webpage footer that lead to pages plausibly relevant to exercising privacy rights. This includes, but is not limited to, links labeled with terms such as: ``Privacy Policy'', ``Your Privacy Rights'', ``Privacy Preferences'', ``Privacy Choices'', ``Consumer Rights''. Do NOT directly click links in footer that labeled ``Do Not Sell My Personal Information'' or other DNSMPI / opt-out--only links, as they are not part of the CCPA Right of Access analysis.\\[2pt]
(Step 2) Navigate to and Review each page obtained from Step 1 across all different aspects as described under [EVALUATION INSTRUCTIONS].\\[2pt]
(Step 3) Enumerate ALL submission methods offered for CCPA Right of Access (webform, email, phone, mail, etc.).\\[2pt]
(Step 4) If a webform is offered as one of the submission methods for CCPA Right of Access, interact with it as a California consumer who exercising Right of Access by selecting the following fields:\\[1pt]
\par\hangindent=0.8cm\hangafter=1\hspace*{0.4cm} - If country is requested, choose or enter ``United States''.\\[1pt]
\par\hangindent=0.8cm\hangafter=1\hspace*{0.4cm} - If state is requested, choose or enter ``California (CA)''.\\[1pt]
\par\hangindent=0.8cm\hangafter=1\hspace*{0.4cm} - If a relationship-to-company field (or similar) is requested, select one. Prefer ``Consumer'' if available.\\[1pt]
\par\hangindent=0.8cm\hangafter=1\hspace*{0.4cm} - If a request type field (or similar) is requested, select the option corresponding to the CCPA Right of Access (or equivalent semantic meaning, e.g., ``Access Data,'' ``Right to Know'').\\[1pt]
\par\hangindent=0.8cm\hangafter=1\hspace*{0.4cm} - If the webform does not display after first navigation, WAIT 5 seconds, then scroll down and up once to trigger lazy-loading, and check again.\\[1pt]
\par\hangindent=0.8cm\hangafter=1\hspace*{0.4cm} - After each input or selection that could change the UI, WAIT 5 seconds for the webform to fully update and reveal any new fields (e.g., the ``request type'' field may only appear after selecting California).\\[2pt]
(Step 5) DO NOT submit form. For each visible form field, record its label and determine if it is a mandatory field or optional. To do this:\\[1pt]
\par\hangindent=0.8cm\hangafter=1\hspace*{0.4cm} - Record mandatory/optional status via UI cues (asterisk, `required' label) and HTML source (`required' attribute).\\[1pt]
\par\hangindent=0.8cm\hangafter=1\hspace*{0.4cm} - Record `unsure' if unclear.
};

\draw[gray!60] (0,-17.8) -- (\W,-17.8);

\end{tikzpicture} 
  }
  \caption{Full agent prompt (Part 1 of 3)}
  \label{fig:prompt_1}
  \Description{Prompt contains role, goal and interaction plan sections.}
\end{figure*}

\begin{figure*}[t]
  \centering
  \resizebox{\textwidth}{!}{
    \usetikzlibrary{positioning}

\definecolor{orangelight}{RGB}{255, 220, 170}
\definecolor{orangeborder}{RGB}{200, 140, 50}
\definecolor{headerbg}{RGB}{242, 242, 242}
\definecolor{greenlabel}{RGB}{180, 230, 180}
\definecolor{yellowlabel}{RGB}{255, 230, 180}
\definecolor{pinklabel}{RGB}{255, 200, 200}
\definecolor{bluelabel}{RGB}{200, 215, 240}
\definecolor{innerboxbg}{RGB}{252, 235, 235}
\definecolor{innerborder}{RGB}{200, 160, 160}
\definecolor{rowtwocolor}{RGB}{255, 240, 200}
\definecolor{rowthreecolor}{RGB}{255, 220, 170}
\definecolor{boxgray}{RGB}{240, 240, 240}
\definecolor{boxgrayborder}{RGB}{180, 180, 180}

\definecolor{colorL1}{HTML}{FFE6CC}
\definecolor{colorL2}{HTML}{FAD7AC}
\definecolor{colorL3}{HTML}{FFD9A0}
\definecolor{colorL4}{HTML}{FFAD5C}

\begin{tikzpicture}[
  every node/.style={font=\small},
  roundtag/.style={
    rounded corners=4pt,
    inner xsep=6pt, inner ysep=3pt, font=\small\bfseries
  },
  smalltag/.style={
    rounded corners=3pt,
    inner xsep=4pt, inner ysep=2pt, font=\footnotesize\bfseries
  },
  labeltag/.style={
    rounded corners=3pt,
    inner xsep=5pt, inner ysep=3pt, font=\footnotesize
  }
]

\def\W{18.5}
\def\TW{17.7cm}

\draw[draw=gray!70, rounded corners=4pt, fill=white]
  (0,0) rectangle (\W, -23.0);

\draw[draw=gray!60, fill=headerbg]
  (0,0) rectangle (\W,-0.85);
\node[anchor=west, font=\small\bfseries] at (0.3,-0.32) {[TAXONOMY]};
\node[labeltag, draw=gray!50, fill=bluelabel, anchor=east]
  at (18.2,-0.32) {Runtime-injected taxonomy};

\draw[gray!40] (0,-0.85) -- (\W,-0.85);

\node[anchor=north west, text width=\TW, font=\small] at (0.3,-0.95) {%
You MUST ONLY use the following dark pattern taxonomy to detect and classify dark patterns: \texttt{\$dark\_pattern\_taxonomy}
};

\node[anchor=north west, text width=\TW, font=\small\itshape] at (0.3,-1.55) {%
(Each dark pattern entry in \texttt{\$dark\_pattern\_taxonomy} contains the following fields, depending on the prompt strategy level used:)
};

\draw[draw=gray!60, rounded corners=3pt, fill=white]
  (0.3,-2.15) rectangle (16.3,-6.7);

\draw[draw=gray!50, fill=white]
  (0.3,-2.15) rectangle (16.3,-2.85);
\node[smalltag, draw=colorL1!80!black, fill=colorL1] at (1.05,-2.5) {L1};
\node[smalltag, draw=colorL2!80!black, fill=colorL2] at (1.65,-2.5) {L2};
\node[anchor=west, font=\small] at (2,-2.5) {Base taxonomy (dark pattern name + definition only)};
\draw[gray!50] (0.3,-2.85) -- (16.3,-2.85);

\draw[draw=gray!50, fill=rowtwocolor]
  (0.3,-2.85) rectangle (16.3,-3.55);
\node[smalltag, draw=colorL3!80!black, fill=colorL3] at (1.05,-3.2) {L3};
\node[anchor=west, font=\small] at (2.0,-3.2) {+ Few shot examples per dark pattern};
\draw[gray!50] (0.3,-3.55) -- (16.3,-3.55);

\draw[draw=gray!50, fill=rowthreecolor]
  (0.3,-3.55) rectangle (16.3,-6.7);
\node[smalltag, draw=colorL4!80!black, fill=colorL4] at (1.05,-4.0) {L4};
\node[anchor=west, font=\small] at (2.0,-4.0) {+ Structured CoT reasoning (evidence with reasoning) inside each few shot example};
\node[anchor=north west, text width=13.5cm, font=\small] at (2.2,-4.4) {%
Evidence with reasoning:\\[1pt]
\hspace*{0.4cm} raw\_snippet \hspace*{1.5cm} Exact text/HTML from page demonstrating the pattern\\[1pt]
\hspace*{0.4cm} action\_taken \hspace*{1.45cm} Specific check or click that revealed it\\[1pt]
\hspace*{0.4cm} dynamic\_outcome \hspace*{0.8cm} What happened after the action\\[1pt]
\hspace*{0.4cm} short\_rationale \hspace*{1.15cm} Why this qualifies as the named dark pattern
};

\draw[gray!60] (0,-6.9) -- (\W,-6.9);

\draw[draw=gray!60, fill=headerbg]
  (0,-6.9) rectangle (\W,-7.55);
\node[anchor=west, font=\small\bfseries] at (0.3,-7.225) {[EVALUATION INSTRUCTIONS]};
\node[labeltag, draw=gray!50, fill=bluelabel, anchor=east]
  at (18.2,-7.225) {Define what to examine};

\draw[gray!40] (0,-7.55) -- (\W,-7.55);

\node[anchor=north west, text width=\TW, font=\small] at (0.3,-7.7) {%
Consider ALL of the following aspects when evaluating clarity, completeness, and accuracy of CCPA Right of Access:\\[1pt]
(1) Static ``policy text'':\\[0pt]
\par\hangindent=0.8cm\hangafter=1\hspace*{0.4cm} - Evaluate the clarity, completeness, and accuracy of how access rights and related instructions are described.\\[0pt]
\par\hangindent=0.8cm\hangafter=1\hspace*{0.4cm} - Identify any omissions, narrowing, ambiguity, fragmentation, inconsistency, or misleading information in access rights workflow or instructions.\\[0pt]
(2) Static/Dynamic ``UI behaviors'':\\[0pt]
\par\hangindent=0.8cm\hangafter=1\hspace*{0.4cm} - Form fields, button labels, color contrasts, hidden links, modal popups, validation messages, redirects, selectors\\[1pt]
(3) Static/Dynamic ``interaction-level obstacles'' encountered during the entire process of exercising Right of Access:\\[0pt]
\par\hangindent=0.8cm\hangafter=1\hspace*{0.4cm} - Availability and validity of privacy-related links\\[0pt]
\par\hangindent=0.8cm\hangafter=1\hspace*{0.4cm} - Issues encountered when following links or completing forms
};

\draw[gray!60] (0,-11.8) -- (\W,-11.8);

\draw[draw=gray!60, fill=headerbg]
  (0,-11.8) rectangle (\W,-12.45);
\node[anchor=west, font=\small\bfseries] at (0.3,-12.125) {[DECISION RULES]};
\node[labeltag, draw=gray!50, fill=bluelabel, anchor=east]
  at (18.2,-12.125) {How to classify dark patterns};

\draw[gray!40] (0,-12.45) -- (\W,-12.45);

\node[anchor=north west, text width=\TW, font=\small] at (0.3,-12.55) {%
Apply the following decision rules when detecting *each* dark pattern during the evaluation of the CCPA Right of Access process, as specified in [EVALUATION INSTRUCTIONS]. Record each observed problem along with its supporting evidence.\\[1pt]
- Exhaustive Coverage: For every dark pattern entry in [TAXONOMY]:\\[0pt]
\par\hangindent=0.8cm\hangafter=1\hspace*{0.4cm} * You MUST generate one \texttt{`DarkPatternMatch`} object for each dark pattern entry and record all findings.\\[0pt]
\par\hangindent=0.8cm\hangafter=1\hspace*{0.4cm} * Match STRICTLY to dark pattern definition's wording. Multiple matches are allowed if and only if multiple entry definitions accurately apply.\\[0pt]
\par\hangindent=0.8cm\hangafter=1\hspace*{0.4cm} * Evaluate independently for every dark pattern entry. Absence or presence must be justified separately.\\[0pt]
\par\hangindent=0.8cm\hangafter=1\hspace*{0.4cm} * ALWAYS consider surrounding and preceding context; do not label ONLY based on a local snippet.\\[1pt]
- Use of Example \textit{(not present in L1/L2)}
};

\draw[draw=boxgrayborder, fill=boxgray, rounded corners=3pt]
  (0.3,-16.4) rectangle (17.8,-18.3);
\node[anchor=north west, text width=15cm, font=\small\itshape] at (1.8,-16.5) {%
(Few-Shot only --- examples field added)
};
\node[smalltag, draw=colorL3!80!black, fill=colorL3] at (1.05,-16.75) {L3};
\node[anchor=west, text width=15cm, font=\small] at (1.8,-17.55) {%
* The `examples' field provides few-shot prototypes showing how each dark pattern may appear in real contexts.\\[0pt]
* Use them to understand the pattern's intent, but do not limit your judgment to exact example phrasings. Be generalizable.
};

\draw[draw=boxgrayborder, fill=boxgray, rounded corners=3pt]
  (0.3,-18.5) rectangle (17.8,-20.0);
\node[anchor=north west, text width=15cm, font=\small\itshape] at (1.8,-18.6) {%
(+ CoT reasoning inside each example )
};
\node[smalltag, draw=colorL4!80!black, fill=colorL4] at (1.05,-18.8) {L4};
\node[anchor=west, text width=13.5cm, font=\small] at (1.8,-19.45) {%
* Each example includes chain-of-thought reasoning under `evidence\_with\_reasoning', including\\[0pt]
raw\_snippet, observed\_ui\_text, action\_taken, dynamic\_outcome, short\_rationale.
};

\node[anchor=north west, text width=\TW, font=\small] at (0.3,-20.15) {%
- Any detected dark pattern MUST have an impact on the user's ability to locate, understand, or complete a CCPA Right of Access request.\\[1pt]
- Content that discusses other rights (e.g., opt-out) may ONLY be cited as evidence if it increases burden, confusion, or distraction specifically within the Right of Access path.\\[1pt]
- Do NOT classify: standard info fields, bot-detection / CAPTCHA measures, hidden DOM elements, unreachable/inferred URLs, or execution failures.
};

\draw[gray!60] (0,-23.0) -- (\W,-23.0);

\end{tikzpicture} 
  }
  \caption{Full agent prompt (Part 2 of 3)}
  \label{fig:prompt_2}
  \Description{Prompt contains taxonomy, evaluation instructions and decision rules sections.}
\end{figure*}

\begin{figure*}[t]
  \centering
  \resizebox{\textwidth}{!}{
    \usetikzlibrary{positioning}

\definecolor{headerbg}{RGB}{242, 242, 242}
\definecolor{bluelabel}{RGB}{200, 215, 240}
\definecolor{greenlabel}{RGB}{180, 230, 180}
\definecolor{yellowlabel}{RGB}{255, 230, 180}
\definecolor{boxgray}{RGB}{240, 240, 240}
\definecolor{boxgrayborder}{RGB}{180, 180, 180}
\definecolor{colorL1}{HTML}{FFE6CC}
\definecolor{colorL2}{HTML}{FFE6CC}
\definecolor{colorL3}{HTML}{FFD9A0}
\definecolor{colorL4}{HTML}{FFAD5C}

\begin{tikzpicture}[
  every node/.style={font=\small},
  smalltag/.style={rounded corners=3pt, inner xsep=4pt, inner ysep=2pt, font=\footnotesize\bfseries},
  labeltag/.style={rounded corners=3pt, inner xsep=5pt, inner ysep=3pt, font=\footnotesize}
]

\def\W{18.5}
\def\TW{17.7cm}

\draw[draw=gray!70, rounded corners=4pt, fill=white] (0,0) rectangle (\W,-20.5);

\draw[draw=gray!60, fill=headerbg] (0,0) rectangle (\W,-0.65);
\node[anchor=west, font=\small\bfseries] at (0.3,-0.325) {[EVIDENCE REQUIREMENTS]};
\draw[gray!40] (0,-0.65) -- (\W,-0.65);

\node[labeltag, draw=gray!50, fill=bluelabel, anchor=east] at (18.2,-0.325)
  {\parbox{4.2cm}{\centering Specify the structured output schema}};
\node[labeltag, draw=gray!50, fill=bluelabel, anchor=east] at (18.2,-2.2)
  {Structured evidence at the pattern level};
\node[labeltag, draw=gray!50, fill=bluelabel, anchor=east] at (18.2,-12.05)
  {Structured evidence for post-run verification};

\node[anchor=north west, text width=\TW, font=\small] at (0.3,-0.8) {%
Populate all fields of the \texttt{`DarkPatternRunResult`} schema.\\[2pt]
(1) Record each discovered/disclosed submission method (e.g., webform, email, phone, mail, etc.)\\[1pt]
(2) Record all visible form fields after fully expanding the UI\\[2pt]
(3) For *each* dark pattern entry:\\[1pt]
\hspace*{0.5cm} (3.1) When a dark pattern is FOUND:\\[0pt]
\hspace*{0.8cm} - Set \texttt{status=``found''}.\\[0pt]
\hspace*{0.8cm} - Provide at least one \texttt{`EvidenceItem`} describing where and how the pattern appears.\\[0pt]
\hspace*{0.8cm} - For each \texttt{`EvidenceItem`}:\\[0pt]
\hspace*{1.2cm} * \texttt{page\_url}: the URL where the evidence was seen.\\[0pt]
\hspace*{1.2cm} * \texttt{raw\_snippet}: copy the exact text or HTML phrase from the page that demonstrates the pattern. Do not paraphrase here.\\[0pt]
\hspace*{1.2cm} * \texttt{observed\_ui\_text}: summarize what that text or behavior implies (why it's a dark pattern).\\[0pt]
\hspace*{1.2cm} * \texttt{action\_taken}: describe the specific check or click that revealed it.\\[0pt]
\hspace*{1.2cm} * \texttt{dynamic\_outcome}: describe what happened (e.g., ``page asked for cookie ID'', ``form disabled until consent'').\\[0pt]
\hspace*{0.8cm} - In \texttt{short\_rationale}, clearly explain why this qualifies as the named dark pattern.\\[0pt]
\par\hangindent=1.2cm\hangafter=1\hspace*{0.8cm} - In \texttt{matching\_confidence}, set between 0.5--1.0 based on clarity of evidence and how confident you sure this dark pattern observed.\\[2pt]
\hspace*{0.5cm} (3.2) When a dark pattern is NOT FOUND\\[0pt]
\hspace*{0.8cm} - Set \texttt{status=``not found''}.\\[0pt]
\hspace*{0.8cm} - Populate evidence with at least one \texttt{`EvidenceItem`} showing where you looked and why no pattern was present.\\[0pt]
\hspace*{0.8cm} - For each \texttt{`EvidenceItem`}:\\[0pt]
\hspace*{1.2cm} * \texttt{page\_url}: the URL you inspected.\\[0pt]
\hspace*{1.2cm} * \texttt{raw\_snippet}: quote the relevant section of the page that led you to conclude no dark pattern exists.\\[0pt]
\hspace*{1.2cm} * \texttt{observed\_ui\_text}: briefly state why this section does not indicate the pattern (e.g., ``form is simple and clearly labeled'').\\[0pt]
\hspace*{1.2cm} * \texttt{action\_taken}: specify what you checked (``reviewed form submission fields'', ``scrolled through privacy section'').\\[0pt]
\hspace*{1.2cm} * \texttt{dynamic\_outcome}: describe normal or compliant behavior (``form submitted successfully'', ``link clearly visible'').\\[0pt]
\hspace*{0.8cm} - In \texttt{short\_rationale}, summarize why this pattern was absent.\\[0pt]
\par\hangindent=1.2cm\hangafter=1\hspace*{0.8cm} - In \texttt{matching\_confidence}, set a score 0.5--1.0 depending on how thoroughly you inspected and how confident you sure this dark pattern do not exist.\\[2pt]
(4) Post-run verification.\\[0pt]
- Rule: A failure should be classified as \texttt{blocked = true} ONLY if it prevents the agent from completing the interaction plan. If the agent encounters a temporary failure, e.g., timeout, but successfully recovers via retry and completes the task, this is NOT considered blocking.\\[1pt]
- If the agent is permanently blocked or unable to fully execute the interaction plan:\\[0pt]
\hspace*{0.8cm} - Choose exactly one \texttt{block\_category}:\\[0pt]
\hspace*{1.2cm} * \texttt{security\_barrier}: website actively blocks automation (CAPTCHA, Cloudflare)\\[0pt]
\hspace*{1.2cm} * \texttt{automation\_instability}: timeout, crash, rendering failure, or infrastructure issue\\[0pt]
\par\hangindent=1.6cm\hangafter=1\hspace*{1.2cm} * \texttt{interaction\_failure}: page loads but cannot interact with UI elements (elements not clickable, dynamic UI not rendered, JS event failure)\\[0pt]
\par\hangindent=1.6cm\hangafter=1\hspace*{1.2cm} * \texttt{navigation\_failure}: required page or section cannot be reached or loaded (broken links, page or section never loads, redirect loops)\\[0pt]
\par\hangindent=1.6cm\hangafter=1\hspace*{1.2cm} * \texttt{content\_format\_limitation}: information exists but cannot be extracted (unreadable PDF, oversized document, text in image, iframe isolation)\\[0pt]
\par\hangindent=1.6cm\hangafter=1\hspace*{1.2cm} * \texttt{agent\_instability}: agent is unable to complete a workflow due to internal failures in execution or planning (incomplete plan execution, wrong output format)\\[0pt]
\hspace*{0.8cm} - set \texttt{blocked} = true\\[0pt]
\hspace*{0.8cm} - include a short \texttt{blocker\_message}\\[0pt]
\hspace*{0.8cm} - \texttt{blocked\_stage} describe where execution failed\\[1pt]
- If no issues occurred, or the agent recovers and completes the task:\\[0pt]
\hspace*{0.8cm} - set \texttt{blocked} = false, set \texttt{block\_category} = ``no\_blocking'', leave other fields null
};

\draw[gray!60] (0,-20.5) -- (\W,-20.5);

\end{tikzpicture} 
  }
  \caption{Full agent prompt (Part 3 of 3)}
  \label{fig:prompt_3}
  \Description{Prompt contains evidence requirements section.}
\end{figure*}

\end{document}